\documentclass[journal]{IEEEtran}

\usepackage{comment}

\usepackage{cite}
\usepackage{amsmath,amssymb,amsfonts}
\usepackage{graphicx}
\usepackage{textcomp}
\usepackage{comment}
\usepackage{cite}
\usepackage{amsmath,amssymb,amsfonts}
\usepackage{algorithmic}
\usepackage{graphicx}
\usepackage{epstopdf}
\usepackage{subfigure}
\usepackage{stfloats}
 \graphicspath{{\figures}}
\usepackage{textcomp}
\usepackage{xcolor}
\usepackage{verbatim}
\usepackage{makecell}
\usepackage{booktabs}
\usepackage{caption}
\usepackage{CJK}
\usepackage{multirow}
\usepackage{caption}

\usepackage{array}
\usepackage{tabularx}
\usepackage{balance}

\usepackage{algorithm}
\usepackage{algorithmic}

\makeatletter
\newcommand{\removelatexerror}{\let\@latex@error\@gobble}
\makeatother

\IEEEoverridecommandlockouts   
\begin{document}
\begin{CJK}{UTF8}{gbsn}

\title{Mobile Embodied Agent Network (MEAN):
Integrated Framework, Enabling Technologies, \\and Task-Driven Deployment}    

\author{%
Zhaoyang Zhang,
Lingxiao Sun,
Weijie Zhou,
Chunhui Feng,
Zhaohui Yang,\\
Xuan Cheng,
Tianyi Qin,
Zhixian Kong,
Baoping Cheng,
Zijian Wu,
and Jingze Che
\thanks{This work has been submitted to the IEEE for possible publication. Copyright may be transferred without notice, after which this version may no longer be accessible. This work was supported in part by National Natural Science Foundation of China under Grants 62631025 and 62394292, and the Fundamental Research Funds for the Central Universities under Grant 226-2024-00069. (\textit{Corresponding author: Zhaoyang Zhang})}
\thanks{Z. Zhang, L. Sun, W. Zhou, C. Feng, Z. Yang, X. Cheng, T. Qin, and Z. Kong are with the College of Information Science and Electronic Engineering, Zhejiang University, Hangzhou 310027, China, and also with the Zhejiang Provincial Laboratory of Multi-Modal Communication Networks and Intelligent Information Processing, Hangzhou 310027, China. Z. Zhang is also with the Institute of Fundamental and Transdisciplinary Research, Zhejiang University, Hangzhou 310058, China. (e-mail: ning\_ming@zju.edu.cn; lingxiao\_sun@zju.edu.cn; wj\_zhou@zju.edu.cn; chunhuifeng@zju.edu.cn; yang\_zhaohui@zju.edu.cn; xuan\_cheng@zju.edu.cn; tianyiqin@zju.edu.cn; zhixian\_kong@zju.edu.cn)}

\thanks{B. Cheng, Z. Wu, and J. Che are with China Mobile (Hangzhou) Information Technology Co., Ltd.,
Hangzhou 311100, China.
(e-mail: chengbaoping@\allowbreak cmhi.chinamobile.com;
wuzijian@\allowbreak cmhi.chinamobile.com;
chejingze@\allowbreak cmhi.chinamobile.com).}
}

\maketitle

\begin{abstract}
Mobile networks are evolving from human-centered, data-oriented communication infrastructures toward machine- and/or agent-dominant, task-oriented platforms that are to serve the rapidly increasing networked machines and embodied agents. In order to fulfill the newly arose complex task objectives, the framework of future mobile network is to be innovated to reflect and facilitate the intrinsic interplay and integration of the underlying processes of perception, communication, computation, decision, action, and feedback, etc. This survey establishes the mobile embodied agent network (MEAN) as a task-oriented networking paradigm in which networking and distributed intelligence are jointly organized around closed-loop embodied task execution. We first establish a unified MEAN framework that explains how network resources and intelligent functions jointly support embodied task execution, and extend its performance indicators from conventional communication metrics to embodied task outcomes and cross-layer system effects. With this framework, we analyze three complementary enabling mechanisms that form a cognition-interaction-adaption loop. Large artificial intelligence (AI) models provide high-level cognition, token communication connects this cognition with distributed agents through shared semantic representations, and reinforcement learning closes the loop by adapting system operation to network, task, and physical feedback. We further investigate the bidirectional interplay between MEAN and emerging technologies, showing how they mutually extend system capabilities and reshape task-oriented network design. Finally, we discuss open challenges and future research directions toward the practical realization of MEAN over real joint communication and computation platform.
\end{abstract}
~\\

\section{Introduction}

The evolution of wireless networks has long been driven by demands for higher data rates, broader connectivity, and more reliable low-latency services. Fifth-generation (5G) and 5G-A systems have advanced these capabilities to support enhanced mobile broadband, massive machine-type communications, and ultra-reliable low-latency communications\cite{itu2015imt2020}. However, the emerging vision of sixth-generation (6G) networks indicates a more fundamental shift: they are expected to move beyond connecting people, devices, and data flows toward supporting connected intelligence in dynamic physical environments\cite{itu2023imt2030,saad2019vision,tataria20216g}. In such a vision, wireless networks will not merely deliver information from one endpoint to another, but will increasingly support integrated sensing, computation, and intelligence to enable closed-loop perception, decision, coordination, and action in applications such as autonomous vehicles, uncrewed aerial vehicle (UAV) swarms, smart manufacturing, and human-robot collaboration\cite{giordani2020toward,dang2020should,mozaffari2019tutorial,cai2025secure}. Unlike passive terminals, the autonomous and mobile agents in these applications continuously interact with the physical environment, making their task execution directly dependent on timely and reliable information exchange, intelligent decision, and environmental feedback.

\begin{figure*}[!t]
    \centering
    \includegraphics[width=1\textwidth]{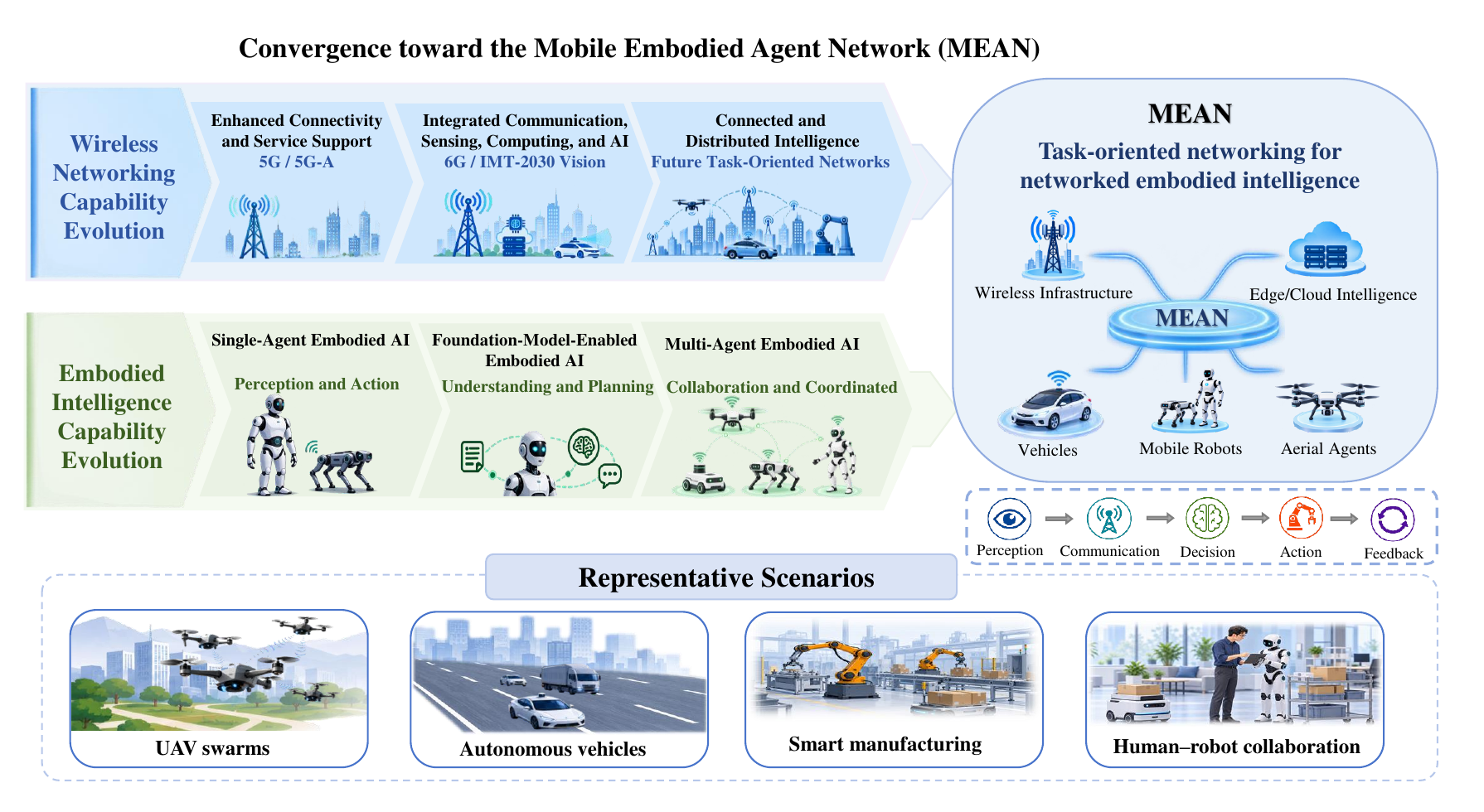}
    \caption{Convergence of wireless networking and embodied intelligence toward the mobile embodied agent network (MEAN).}
    \label{fig:MEAN_evolution}
\end{figure*}

Embodied artificial intelligence serves as the key driving force in this transition. Unlike conventional AI systems that mainly operate in digital spaces with static datasets or one-shot inference pipelines, embodied agents perceive the physical world through sensors, reason over environmental states and task objectives, and execute actions through physical bodies or actuators\cite{savva2019habitat,kolve2017ai2,shridhar2020alfworld,srivastava2022behavior}. Recent advances in embodied multimodal models, vision-language-action models, and robotic foundation models have further enhanced their capabilities for semantic understanding, task planning, and adaptive physical interaction\cite{driess2023palm,zitkovich2023rt,o2024open}. Nevertheless, future embodied intelligence cannot rely solely on isolated single-agent systems. In many practical scenarios, embodied agents are spatially distributed, resource-constrained, and required to cooperate under changing network and physical conditions. Their task performance is therefore determined not only by the intelligence of individual agents, but also by the quality of networked collaboration among agents, embodied platforms, edge infrastructures, and cloud intelligence. This trend motivates a shift from embodied AI as an individual capability to networked embodied intelligence as a system-level capability, where the network must coordinate not only data flows and resources, but also models, contexts, actions, and feedback across distributed agents. As illustrated in Fig.~\ref{fig:MEAN_evolution}, the evolution of embodied intelligence and wireless networking is converging toward a unified system paradigm for networked embodied intelligence.

Advances in integrated wireless networking and edge intelligence provide connectivity and distributed computing support for embodied agents. Embodied AI complements these capabilities with physical perception and task reasoning. However, these capabilities still need to be integrated within a unified, task-oriented networking framework. Mobile embodied agents continuously use observations and execution feedback to guide subsequent actions. Supporting this closed loop requires wireless communication and distributed intelligence to be coordinated around task progress and interactions among agents. To address this need, we introduce the concept of mobile embodied agent network (MEAN), which provides a unified networking framework for connecting and coordinating mobile embodied agents, wireless infrastructures, distributed intelligence, and physical environments to support closed-loop task execution. The specific research gaps and positioning of MEAN are discussed in the following subsection.

\subsection{Motivation}

Existing wireless-AI research provides complementary foundations for intelligent mobile systems. Integrated sensing and communication (ISAC) coordinates environmental sensing with information transmission, while integrated sensing, communication, and computation (ISCC) extends this joint design to information processing\cite{liu2022survey,wen2024survey}. Mobile edge computing and edge intelligence bring computing resources and AI inference closer to mobile agents\cite{mach2017mec,porambage2018mec,park2019wireless}. At the network level, AI-radio access network (AI-RAN) introduces AI-driven optimization, while open-radio access network (O-RAN) provides programmable interfaces for network control\cite{polese2023understanding}. Together, these capabilities allow agents to draw on networked resources for perception and reasoning beyond what individual devices can support.

\begin{figure*}[!t]
    \centering
    \includegraphics[width=0.95\textwidth]{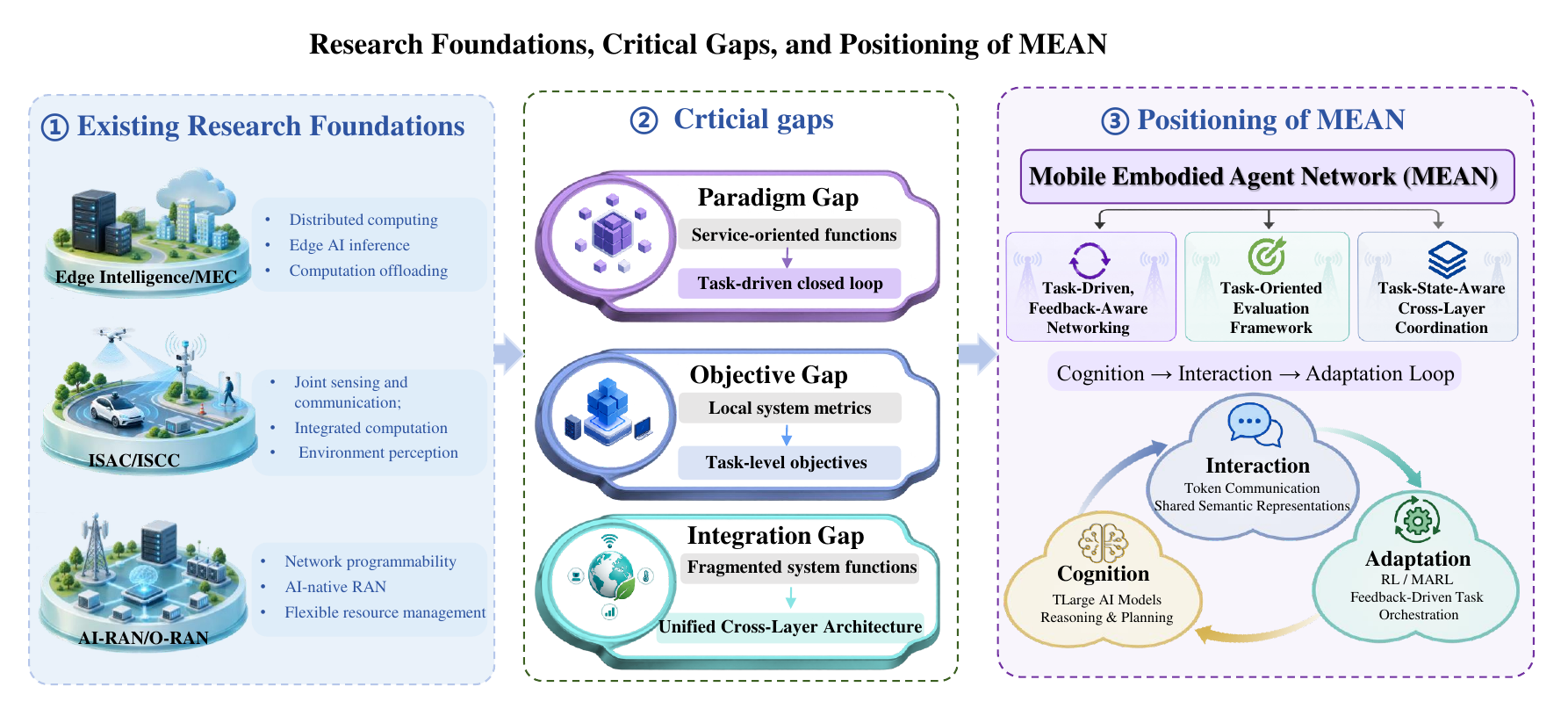}
    \caption{Research foundations, critical gaps, and positioning of MEAN for networked embodied intelligence.}
    \label{fig:motivation_positioning}
\end{figure*}

However, these capabilities alone do not ensure coherent support for closed-loop embodied tasks. Network decisions determine whether agents receive the information and computing support needed to act in time, while their actions change the physical environment and subsequent network demands. Supporting embodied intelligence therefore requires network operation and physical task execution to adapt to each other through continuous feedback. As summarized in Fig.~\ref{fig:motivation_positioning}, three gaps separate existing wireless-AI designs from this requirement, concerning the organizing paradigm, optimization objectives, and cross-layer integration.

\textbf{Paradigm gap.}
Existing wireless systems are largely organized around delivering communication and computing services, with service completion often treated as the endpoint of network operation. For embodied agents, however, delivering data or an inference result is only an intermediate step toward completing a physical task. Its value depends on whether it enables the agent to take an appropriate action at the required time. Delayed information, for example, may describe a state that has already changed and lead to an unsuitable action. The resulting physical outcome must therefore inform subsequent network decisions. This coupling calls for a task-driven, feedback-aware networking paradigm that supports the full course of embodied task execution.

\textbf{Objective gap.}
Conventional evaluation emphasizes subsystem-level metrics such as communication latency, inference accuracy, and energy consumption, whereas embodied task success depends on how these factors jointly affect physical execution. Optimizing these metrics independently may overlook their interactions, so improvements in individual metrics do not necessarily lead to more timely, reliable, or successful task completion. For example, a more accurate model may require additional processing time, making its output less useful for an action that must be taken promptly. MEAN therefore requires an evaluation framework that relates basic system performance to embodied task outcomes and captures cross-layer intelligence and resource tradeoffs. Representative measures include task success rate, mission completion time, and safe success rate for task outcomes, together with planning validity and task–resource efficiency for cross-layer evaluation\cite{gunduz2022beyond,garcia2015comprehensive,xu2023edge}.

\textbf{Integration gap.}
The technical foundations required by MEAN remain fragmented across system layers. Communication, sensing, computation, intelligence, and control are often designed with different interfaces, timescales, and optimization objectives, although decisions in one layer can directly affect the operating conditions of others. Agent mobility and environmental changes alter communication and sensing conditions, while resource allocation and computation placement affect inference latency, coordination, and subsequent physical actions. Addressing these dependencies requires a shared representation of task state and cross-layer coordination mechanisms that align resource decisions with the needs of physical execution. 

Addressing these gaps requires complementary capabilities for cognition, interaction, and adaptation. Large AI models provide the cognitive foundation by interpreting multimodal observations and task instructions to support reasoning and planning\cite{yao2022react,schick2023toolformer}. Realizing these capabilities across distributed agents requires task-relevant information to be exchanged and aligned. Token communication supports this interaction through shared semantic representations, enabling agents to exchange context and coordinate their contributions to a common task\cite{yang2022semantic,lu2023semantics,liu2024survey}. The resulting plans must also adapt to changing physical environments. RL and multi-agent reinforcement learning (MARL) support this adaptation by using network states and execution feedback to adjust task orchestration\cite{sutton1998reinforcement,mnih2015human,luong2019applications,feriani2021single,hernandez2019survey}.

Together, these technologies form a cognition-interaction-adaptation loop. Model-based reasoning guides what agents need to communicate and accomplish, token exchanges provide shared context for collaborative execution, and feedback-driven adaptation adjusts how network resources support reasoning and communication. These adjustments shape the information and resources available for subsequent reasoning and interaction. MEAN provides a unified networking framework for integrating this loop into distributed embodied task execution. The following subsection formalizes MEAN and clarifies its core capabilities and positioning.

\subsection{Definition of MEAN}

In this survey, MEAN refers to a task-oriented wireless networking paradigm in which mobile embodied agents, wireless infrastructures, cloud-edge-device intelligence, and physical environments are jointly organized to support networked embodied task execution. Unlike conventional wireless systems centered on data delivery or isolated intelligent services, MEAN coordinates networking and distributed intelligence according to task requirements and physical-world feedback.

MEAN is characterized by four key features:
\begin{itemize}
    \item \textbf{Physical closed-loop operation.}
    Network operation is coupled with perception, decision, action, and physical feedback during task execution.

    \item \textbf{Semantic interaction.}
    Agents exchange task-relevant information, contextual states, and coordination information rather than relying solely on raw data transmission.

    \item \textbf{Task-driven deployment.}
    Models, computing functions, and network resources are deployed and adapted across cloud, edge, and devices according to task requirements and system conditions.

    \item \textbf{Multi-agent coordination.}
    Distributed embodied agents and network infrastructures coordinate information, resources, and actions toward common or interdependent task objectives.
\end{itemize}

To further clarify the scope and positioning of MEAN, Table~\ref{tab:MEAN_positioning} compares MEAN with several representative research directions that provide important foundations for networked embodied intelligence, including ISAC/ISCC\cite{liu2022survey,wen2024survey}, edge intelligence and AI-RAN\cite{mach2017mec,porambage2018mec,park2019wireless,polese2023understanding}, semantic communication
\cite{yang2022semantic,lu2023semantics,liu2024survey}, and embodied AI\cite{savva2019habitat,kolve2017ai2,shridhar2020alfworld,srivastava2022behavior}. The comparison focuses on their primary research objectives and their typical emphasis on physical closed-loop operation, semantic interaction, task-driven deployment, and multi-agent coordination.

As shown in Table~\ref{tab:MEAN_positioning}, these research directions provide complementary foundations for MEAN while emphasizing different aspects of networked intelligence. ISAC and ISCC primarily integrate communication with sensing and computation, whereas edge intelligence and AI-RAN provide distributed computing, intelligence, and programmable network control. Semantic communication emphasizes task-relevant information exchange, while embodied AI focuses primarily on perception, reasoning, planning, and physical interaction at the agent level. MEAN builds on these foundations from a system-level perspective by jointly considering physical closed-loop operation, semantic interaction, task-driven deployment, and multi-agent coordination for networked embodied task execution. Accordingly, this survey focuses on the intersection of wireless networking, embodied intelligence, semantic interaction, distributed intelligence, and physical world task execution. We organize the survey around the system architecture of MEAN, its enabling technologies, and task-driven deployment, and further discuss the associated design challenges and open research directions.

\begin{table*}[!t]
\centering
\caption{Positioning of MEAN Relative to Related Research Directions}
\label{tab:MEAN_positioning}

\small
\renewcommand{\arraystretch}{1.15}
\setlength{\tabcolsep}{4pt}
\renewcommand{\tabularxcolumn}[1]{m{#1}}

\begin{tabularx}{\textwidth}{
    @{}
    c
    >{\raggedright\arraybackslash}X
    c c c c
    @{}
}
\toprule
\textbf{Direction}
&
\textbf{Primary Focus}
&
\makecell[c]{\textbf{Physical Closed-Loop}\\\textbf{Operation}}
&
\makecell[c]{\textbf{Semantic}\\\textbf{Interaction}}
&
\makecell[c]{\textbf{Task-Driven}\\\textbf{Deployment}}
&
\makecell[c]{\textbf{Multi-Agent}\\\textbf{Coordination}}
\\
\midrule

ISAC \cite{liu2022survey}
& Joint sensing and communication
& Partial & Limited & Limited & Limited
\\

ISCC \cite{wen2024survey}
& Joint sensing, communication, and computation
& Partial & Limited & Partial & Partial
\\

Edge Intelligence \cite{mach2017mec,park2019wireless}
& Distributed computation, AI inference, and edge intelligence
& Partial & Partial & Partial & Partial
\\

AI-RAN \cite{khani2024airan}
& AI/ML-enabled RAN optimization and resource management
& Partial & Limited & Partial & Partial
\\

Semantic Communication \cite{yang2022semantic,lu2023semantics}
& Task-relevant and meaning-oriented information transmission
& Partial & Strong & Limited & Partial
\\

Embodied AI \cite{savva2019habitat,driess2023palm,zitkovich2023rt}
& Perception, reasoning, planning, and action of embodied agents
& Strong & Partial & Limited & Partial
\\

\textbf{MEAN}
& \textbf{Networked embodied task execution}
& \textbf{Strong}
& \textbf{Strong}
& \textbf{Strong}
& \textbf{Strong}
\\

\bottomrule
\end{tabularx}

\par\smallskip
\begin{minipage}{\textwidth}
\footnotesize
``Strong,'' ``Partial,'' and ``Limited'' indicate the typical research emphasis
identified from representative literature rather than the absence of the
corresponding capability.
\end{minipage}

\end{table*}

\subsection{Contributions}

The main contributions of this survey are summarized as follows.

\begin{itemize}

\item We develop a system architecture and evaluation framework for MEAN. The architecture organizes mobile embodied agents, wireless networks, edge--cloud resources, closed-loop functional capabilities, and cross-layer intelligence within a common system view. The evaluation framework further categorizes performance measures into basic system, embodied task, and cross-layer intelligence and tradeoff indicators.

\item We provide a systematic review of three enabling technologies for MEAN, including reinforcement learning for feedback-driven adaptation, token communication for distributed semantic interaction, and large AI models for cognition and task-driven deployment. We examine how their complementary roles form a cognition-interaction-adaptation loop that connects reasoning, information exchange, and adaptive orchestration with embodied task execution.

\item We synthesize the interplay between MEAN and emerging technologies, including non-terrestrial networks, world models, digital twins, integrated sensing and communication, and edge intelligence. We examine how these technologies extend MEAN capabilities and how embodied task requirements, agent states, and execution feedback reshape their system design.

\item We identify open research challenges and future directions for MEAN, covering cross-layer co-design, reliable and efficient token-mediated interaction, scalable multi-agent coordination, real-time deployment of large AI models, safety and trustworthiness, and benchmarking and evaluation.

\end{itemize}

\subsection{Organization}

\begin{figure*}[t]
    \centering
    \includegraphics[width=0.95\textwidth]{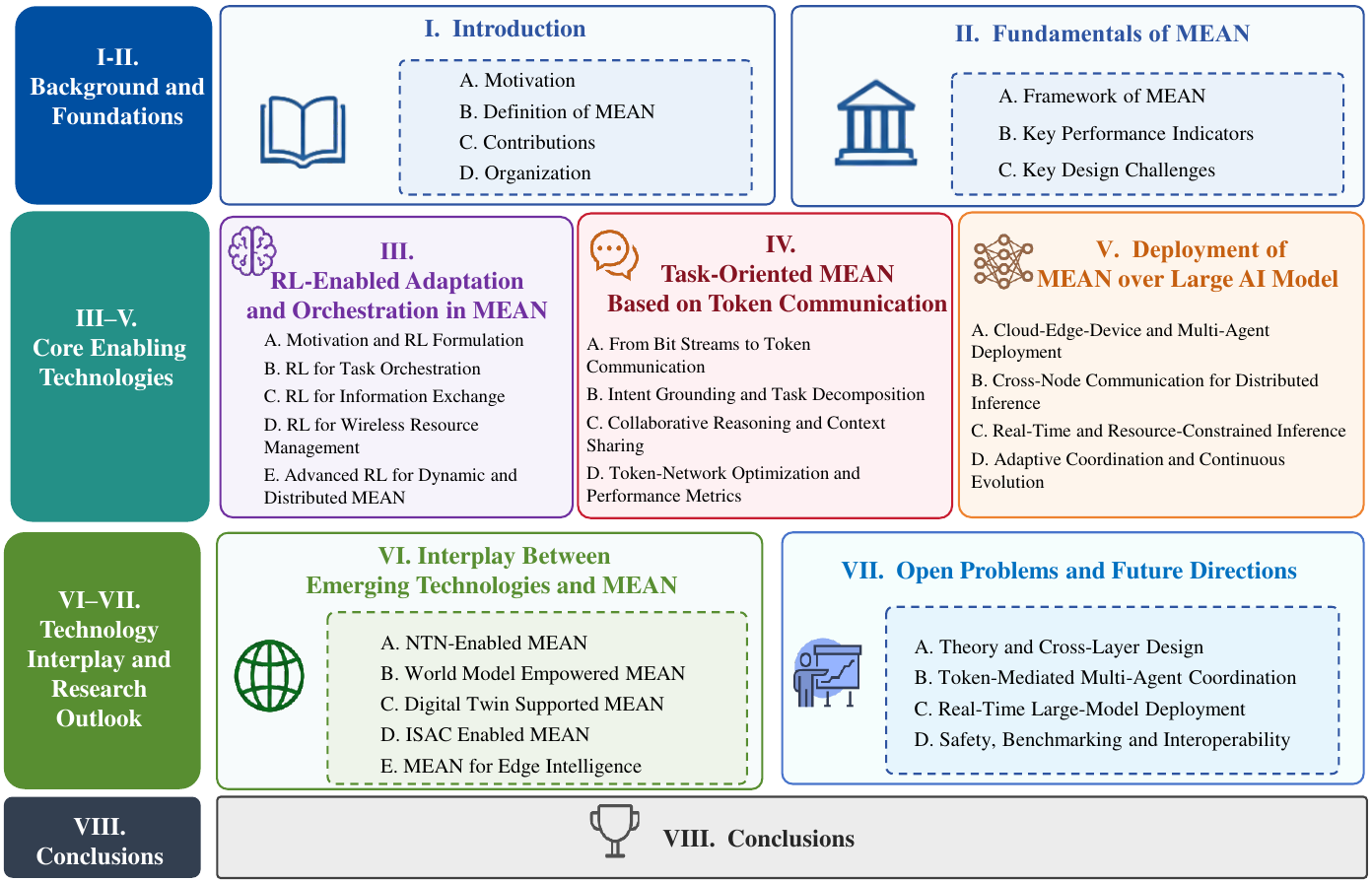}
    \caption{Structure of this survey. }
    \label{fig:survey_structure}
\end{figure*}

The remainder of this survey is organized as follows, as illustrated in Fig.~\ref{fig:survey_structure}. Section II introduces MEAN fundamentals, including its framework, performance indicators, and design challenges. Section III reviews RL-enabled task orchestration, information exchange, and resource management. Section IV examines token-based semantic representation, task planning, collaborative reasoning, and network optimization. Section V discusses large AI model deployment and adaptive inference under resource and network constraints. Section VI explores connections with non-terrestrial networks, world models, digital twins, integrated sensing and communication, and edge intelligence. Section VII outlines open problems and future research directions. Finally, Section VIII concludes the survey.

\section{Fundamentals of MEAN}
This section presents the fundamental architecture of MEAN, including its layered framework, evaluation indicators, and system-level advantages and challenges.

\subsection{Framework of MEAN}
\begin{figure*}[t]
    \centering
    \captionsetup{justification=justified,singlelinecheck=false}
    \includegraphics[width=0.95\linewidth]{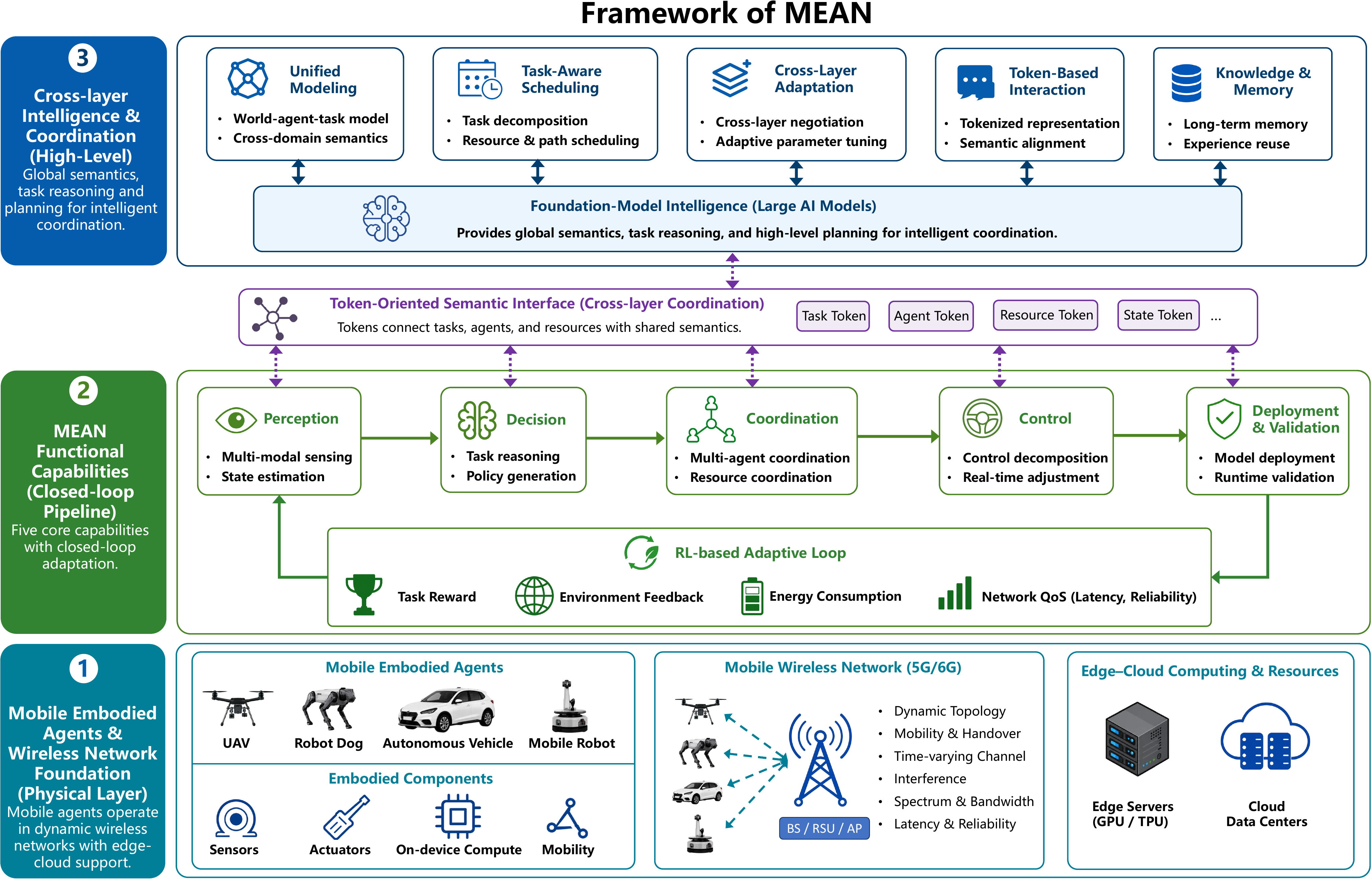}
    \caption{The integrated framework of MEAN. MEAN integrates mobile embodied agents, wireless networking, and edge–cloud resources with a closed-loop pipeline of perception, decision, coordination, control, and deployment and validation. }
    \label{fig:MEAN_framework}
\end{figure*}

As shown in Fig.~\ref{fig:MEAN_framework}, MEAN can be organized into three tightly coupled layers: the mobile embodied agent and wireless network foundation, the closed-loop functional capability pipeline, and the high-level cross-layer intelligence and coordination layer. These layers are connected through a token-oriented semantic interface and an RL-based adaptive loop, enabling task-oriented closed-loop intelligence across agents, wireless networks, computing resources, and intelligent coordination modules.

The mobile embodied agent and wireless network foundation provides the physical and infrastructural substrate of MEAN. It consists of mobile embodied agents, mobile wireless networks, and edge--cloud computing resources. Mobile agents, such as UAVs, robot dogs, autonomous vehicles, and mobile robots, integrate sensors, actuators, on-device computing, and mobility to perceive and interact with the physical environment \cite{savva2019habitat,kolve2017ai2thor,shridhar2020alfworld,srivastava2022behavior}. Meanwhile, 5G/6G wireless networks connect mobile agents with BSs, RSUs, APs, and edge--cloud infrastructure, while facing mobility, handover, time-varying channels, interference, spectrum constraints, and latency and reliability requirements. Edge servers and cloud data centers further provide heterogeneous computation and storage resources \cite{mach2017mec,porambage2018mec,liu2019edge,wen2024iscc}. This layer therefore jointly supports physical interaction, wireless connectivity, and distributed computation.

The closed-loop functional pipeline transforms task intelligence into concrete embodied behavior through five core capabilities: perception, decision, coordination, control, and deployment and validation. Perception extracts task-relevant states from multimodal observations; decision performs task reasoning and policy generation; coordination supports multi-agent cooperation and resource coordination; control decomposes high-level decisions into executable commands with real-time adjustment; and deployment and validation maps models and policies onto actual systems and evaluates their runtime performance \cite{savva2019habitat,srivastava2022behavior,ahn2023saycan,driess2023palme}. These capabilities form a closed loop in which execution outcomes continuously influence subsequent perception, decision, coordination, and control.

The high-level cross-layer intelligence and coordination layer provides system-level orchestration for MEAN. It supports unified modeling, task-aware scheduling, cross-layer adaptation, token-based interaction, and knowledge and memory. Unified modeling captures world--agent--task relationships and cross-domain semantics; task-aware scheduling performs task decomposition and resource/path scheduling; cross-layer adaptation dynamically adjusts system parameters across layers; token-based interaction provides semantic representations for aligning tasks, agents, resources, and states \cite{yang2023semantic,lu2024semantics,liu2024semantic,trevlakis2024semantic}; and knowledge and memory enable long-term context preservation and experience reuse. Foundation-model intelligence further provides global semantics, task reasoning, and high-level planning for intelligent coordination \cite{kim2024llmrobotics,driess2023palme,zitkovich2023rt2}.

Crucially, two cross-layer mechanisms tightly couple the three layers. First, the \emph{token-oriented semantic interface} establishes a unified representation of task, agent, resource, and state tokens, enabling semantic information exchange and coordinated interaction across heterogeneous agents, network infrastructure, and intelligence modules. By abstracting heterogeneous information into task-relevant tokens, this interface bridges physical observations, network resources, and high-level reasoning.

Second, the \emph{RL-based adaptive loop} enables task-driven cross-layer optimization by incorporating task rewards, environmental feedback, energy consumption, and network QoS, including latency and reliability. Through continuous interaction with the environment, RL dynamically adjusts communication, computation, coordination, and control policies according to task requirements and network conditions, thereby connecting system-level resource adaptation with embodied task execution.

\subsection{Key Indicators}

Evaluating MEAN is inherently more complicated than evaluating a conventional communication system or an isolated AI model. Since MEAN supports embodied task execution over dynamic wireless networks, its performance should be assessed not only by communication and computation efficiency, but also by whether the overall closed loop can sustain perception, decision, coordination, control, and physical interaction. Therefore, as summarized in Table~\ref{tab:MEAN_indicators}, the key indicators of MEAN can be organized into three levels: basic system indicators, embodied task indicators, and cross-layer intelligence and tradeoff indicators \cite{mach2017mec,wen2024iscc,yang2023semantic,srivastava2022behavior}.

\begin{table*}[!t]
\centering
\caption{Three-Level Indicator System for MEAN Evaluation}
\label{tab:MEAN_indicators}

\small
\renewcommand{\arraystretch}{1.15}
\setlength{\tabcolsep}{4pt}
\renewcommand{\tabularxcolumn}[1]{m{#1}}

\begin{tabularx}{\textwidth}{
    @{}
    >{\centering\arraybackslash}m{0.16\textwidth}
    >{\centering\arraybackslash}m{0.19\textwidth}
    >{\raggedright\arraybackslash}X
    @{}
}
\toprule
\textbf{Level}
& \textbf{Category}
& \multicolumn{1}{c}{\textbf{Representative Metrics}}
\\
\midrule

\multirow[c]{2}{*}{\textbf{Basic System}}
& Communication
& Latency, Reliability, Throughput, Jitter,
  Packet Delivery Ratio (PDR), Age of Information (AoI)
  \cite{mach2017mec,wen2024iscc,yang2023semantic}
\\
& Computation
& Inference Latency, Time to First Token (TTFT),
  Time per Output Token (TPOT), Token Throughput,
  Energy Consumption, Offloading Overhead
  \cite{kwon2023efficient,zhong2024distserve}
\\
\midrule

\multirow[c]{2}{*}{\textbf{Embodied Task}}
& Task Effectiveness
& Task Success Rate, Goal-Condition Success,
  Task Progress, Mission Completion Time
  \cite{shridhar2020alfworld,srivastava2022behavior,yang2025embodiedbench}
\\
& Embodiment \& Safety
& Motion Smoothness, Control Stability, Collision Rate,
  Safe Success Rate, Constraint Violation Rate
  \cite{garcia2015saferl,ahn2023saycan}
\\
\midrule

\multirow[c]{3}{*}{
    \makecell[c]{
        \textbf{Cross-Layer}\\
        \textbf{Intelligence}\\
        \textbf{\& Tradeoff}
    }
}
& System Scalability
& Scalability, Interoperability, Topology Resilience,
  Coordination Robustness, Multi-Agent Efficiency
  \cite{yang2023semantic,lu2024semantics,liu2024semantic}
\\
& Tradeoff Efficiency
& Task--Resource Efficiency,
  Latency--Energy--Accuracy Tradeoff,
  Coordination Overhead,
  Communication--Computation--Control Balance
  \cite{wen2024iscc,trevlakis2024semantic}
\\
& \makecell[c]{Foundation-Model\\Intelligence}
& Instruction-Following Accuracy, Reasoning Accuracy,
  Planning Validity, Semantic Grounding Accuracy,
  Hallucination Rate, Token Efficiency
  \cite{driess2023palme,zitkovich2023rt2,kim2024llmrobotics}
\\

\bottomrule
\end{tabularx}
\end{table*}

\textbf{Basic system indicators:}
The first level evaluates the communication and computation capabilities required to support embodied intelligence. Communication metrics include latency, reliability, throughput, jitter, packet delivery ratio (PDR), and age of information (AoI). The end-to-end latency can be written as $T_{\mathrm{e2e}} = T_{\mathrm{tx}}+T_{\mathrm{queue}}+T_{\mathrm{proc}}+T_{\mathrm{prop}}$, where the four terms represent transmission, queuing, processing, and propagation delays, respectively. These metrics directly affect control responsiveness, semantic consistency, and multi-agent coordination \cite{yang2023semantic,lu2024semantics}.

Computation indicators include inference latency, TTFT, TPOT, token throughput, energy consumption, and offloading overhead, reflecting whether distributed and large-model-supported intelligence can operate under real-time and resource constraints \cite{mach2017mec,kwon2023efficient,zhong2024distserve}. In particular, TTFT and TPOT characterize the response and generation latency of large models, while token throughput reflects their serving efficiency.

\textbf{Embodied task indicators:}
The second level evaluates whether the embodied mission can be completed effectively and safely. General task-level metrics include task success rate, goal-condition success, task progress, and mission completion time. Goal-condition success can be expressed as $\mathrm{GCS}=\frac{1}{N}\sum_{i=1}^{N}\frac{|\mathcal{G}^{\mathrm{done}}_i|}{|\mathcal{G}_i|}$, where \(\mathcal{G}_i\) denotes the required goal conditions and \(\mathcal{G}^{\mathrm{done}}_i\) denotes the completed subset. This metric captures partial task accomplishment even when the overall mission is not fully completed \cite{shridhar2020alfworld,srivastava2022behavior}.

Safety and embodiment indicators include motion smoothness, control stability, collision rate, safe success rate, and constraint violation rate. A representative metric is the safe success rate, $\mathrm{SSR}=\frac{1}{N}\sum_{i=1}^{N}\mathbb{I}(S_i=1)\mathbb{I}(V_i=0)$, where \(S_i=1\) indicates successful task completion and \(V_i=0\) indicates that no safety or constraint violation occurs. These indicators capture whether MEAN can accomplish embodied tasks without sacrificing physical safety and execution reliability \cite{garcia2015saferl,ahn2023saycan}.

\textbf{Cross-layer intelligence and tradeoff indicators:}
The third level evaluates how effectively MEAN coordinates communication, computation, control, task execution, and large-model intelligence. Scalability, interoperability, topology resilience, coordination robustness, and multi-agent support efficiency characterize the system's ability to operate as the number of agents and the complexity of interactions increase.

Cross-layer efficiency further measures task performance under limited communication and computation resources. A representative task--resource efficiency metric can be defined as $\mathrm{TRE}=\frac{\mathrm{TaskScore}}{\alpha \bar{T}+\beta \bar{E}+\gamma \bar{B}+\delta \bar{C}}$, where \(\bar{T}\), \(\bar{E}\), \(\bar{B}\), and \(\bar{C}\) denote normalized latency, energy consumption, communication overhead, and computational cost, respectively. This metric reflects the fundamental tradeoff between task performance and resource consumption.

For foundation-model-enhanced intelligence, relevant indicators include instruction-following accuracy, reasoning accuracy, planning validity, semantic grounding accuracy, hallucination rate, and token efficiency. These metrics characterize whether large models improve semantic reasoning and coordination without introducing excessive computation, communication, or token overhead.

\subsection{Key Design Challenges}

Realizing MEAN requires addressing three fundamental system-level challenges arising from dynamic environments, constrained wireless communication, and limited computational resources.

\textit{1) Task and Resource Coordination in Dynamic Environments:}
Mobile agents, wireless conditions, and computing resources evolve dynamically, making it challenging to coordinate task execution and resource allocation across heterogeneous components. MEAN requires adaptive cross-layer mechanisms to jointly optimize task decisions, communication, computation, and control while maintaining closed-loop stability. Section~III explores RL-based adaptation to address this challenge \cite{busoniu2008marl,ning2024marl}.

\textit{2) Information Effectiveness and Consistency under Constrained Communication:}
Limited bandwidth, variable latency, and unreliable wireless links constrain information exchange among agents and distributed intelligence modules. MEAN must preserve task-relevant information and semantic consistency while minimizing communication overhead, particularly under compression and transmission errors. Section~IV investigates token-oriented communication to support efficient and reliable semantic interaction \cite{yang2023semantic,lu2024semantics}.

\textit{3) Real-Time and Reliable Intelligence under Resource Constraints:}
Deploying advanced AI models on mobile and edge platforms is constrained by limited computation, memory, and energy resources. MEAN must balance reasoning and planning capabilities with inference latency, resource efficiency, and execution reliability, while mitigating uncertain or unsafe model outputs. Section~V examines efficient and trustworthy deployment of large AI models for embodied intelligence \cite{driess2023palme,zitkovich2023rt2,kim2024llmrobotics}.

\section{RL-Enabled Adaptation and Orchestration in MEAN}\label{RL-enabled MEAN}

\subsection{Motivation and RL Formulation}

MEAN integrates mobile embodied agents, wireless networks, and distributed computing infrastructures to support closed-loop perception, reasoning, communication, and physical interaction. Unlike conventional mobile networks that primarily transport data between endpoints, MEAN directly supports task execution in dynamic physical environments. Mobile robots, UAVs, autonomous vehicles, and other embodied agents continuously perceive their surroundings, exchange task-relevant information, invoke local or edge intelligence, coordinate with other agents, and adjust their physical actions according to environmental feedback. Consequently, the effectiveness of MEAN depends not only on communication efficiency, but also on how communication, computing, intelligence, and agent capabilities are dynamically orchestrated according to evolving task requirements.

Such orchestration is particularly challenging because the operating conditions of MEAN vary across both the physical and cyber domains. Agent mobility changes network topology and channel quality; sensing uncertainty alters the value of newly collected observations; task phases introduce different latency, reliability, and computation requirements; and limited spectrum, energy, and computing resources constrain the set of feasible decisions. Moreover, these factors are tightly coupled. For example, transmitting high-resolution sensory information may improve perception quality but consume substantial wireless resources, whereas feature compression can reduce communication overhead at the cost of information loss. Offloading an inference task to an edge server may improve computation capability but introduce transmission and queueing delays. Similarly, prioritizing one safety-critical agent may temporarily reduce the resources available to other agents while improving overall mission reliability. Therefore, MEAN requires an adaptive mechanism that can continuously balance task utility against communication, computation, energy, and safety costs.

\begin{figure*}[!t]
\centering
\includegraphics[width=0.95\textwidth]{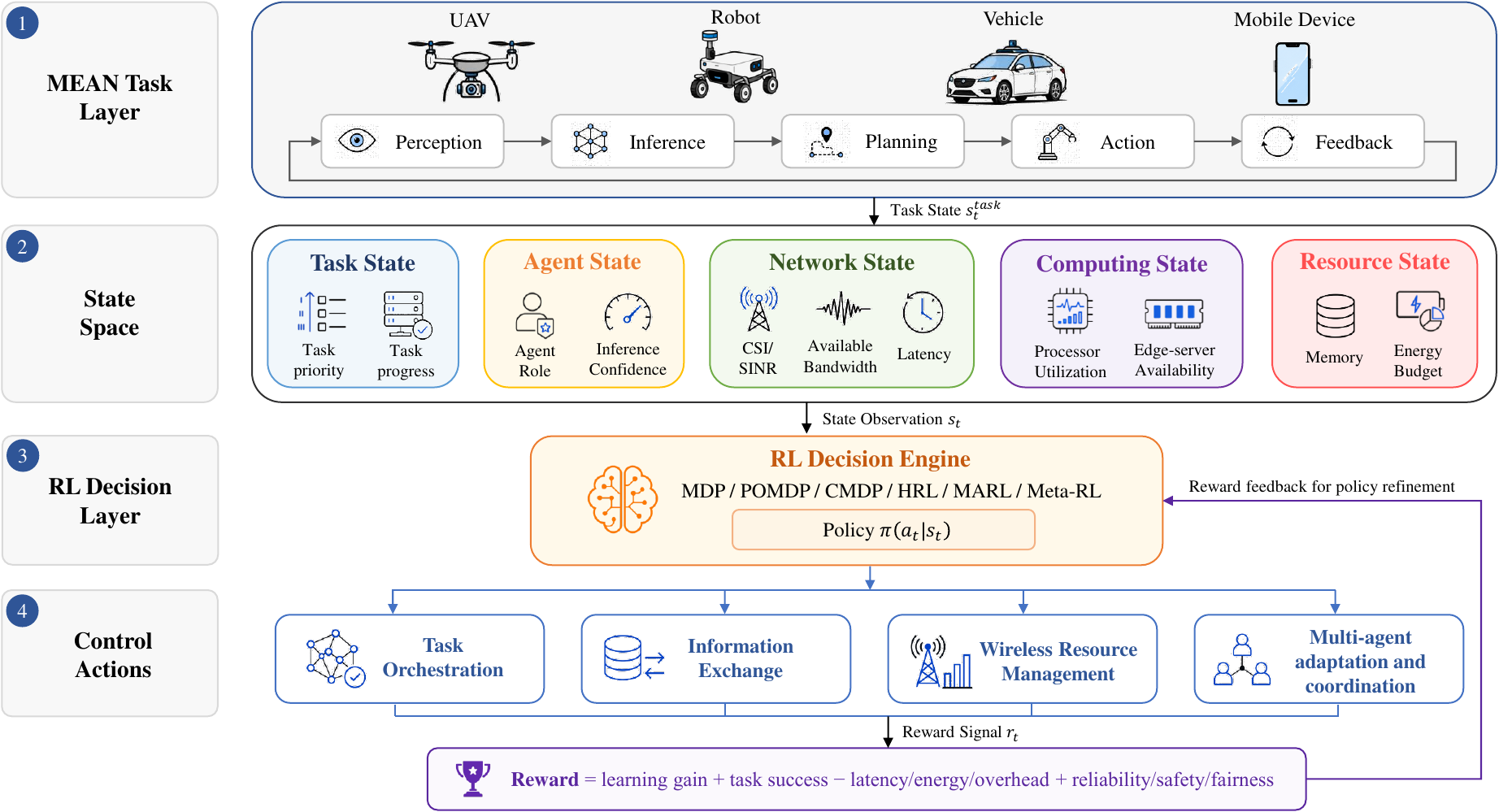}
\caption{RL-enabled adaptation and orchestration framework for MEAN. Mobile embodied agents interact with physical and network environments and generate task, agent, network, computing, and resource states. The RL decision engine maps these multidimensional states to task orchestration, task-aware information exchange, network resource management, and multi-agent adaptation and coordination. Task execution and network operation provide feedback on task utility, latency, reliability, communication and energy costs, safety, and fairness for policy adaptation.}
\label{fig:rl_mean_framework}
\end{figure*}

This problem is inherently sequential. A decision made at the current time may affect not only instantaneous network performance but also future task progress and system states\cite{bai2023toward,zangooei2023reinforcement}. Allocating additional bandwidth to a perception stream may improve subsequent localization and planning; postponing a non-critical model update may save resources in the current stage but reduce model freshness later; selecting edge-assisted inference may improve decision quality while changing future communication load; and modifying the cooperation pattern among agents may reshape both information availability and resource competition. These delayed and coupled effects make static rules and one-shot optimization insufficient for highly dynamic MEAN deployments.

Reinforcement learning (RL) provides a natural framework for addressing such task-driven adaptation and orchestration. By continuously interacting with the environment, an RL policy can learn how current decisions influence long-term task outcomes under uncertain and time-varying conditions. Rather than optimizing a single instantaneous metric, such as throughput, latency, or inference accuracy, RL can jointly account for task completion, information value, communication efficiency, computation cost, energy consumption, reliability, and physical safety\cite{sutton1998reinforcement,mnih2015human}. More importantly, RL allows orchestration policies to adapt according to task phase, agent state, mobility, network conditions, and accumulated execution feedback, making it well suited for the closed-loop operation of MEAN.

To provide a unified view, Fig.~\ref{fig:rl_mean_framework} illustrates an RL-enabled adaptation and orchestration framework for MEAN. At each decision epoch, the MEAN environment produces a multidimensional system state reflecting ongoing embodied tasks and available system resources. The task and embodied-agent state may include task progress, task priority, agent role, mobility, sensing quality, inference confidence, model freshness, and safety relevance. The network state characterizes channel quality, bandwidth availability, interference, queue status, latency, reliability, and network topology. The computing and resource state further captures processor utilization, edge-server availability, memory, energy budget, and other resource constraints. By jointly observing these states, the RL decision engine learns an orchestration policy that dynamically maps system conditions to task- and resource-level actions.

The resulting RL actions span several tightly coupled control dimensions. First, task orchestration determines how embodied tasks are decomposed, prioritized, assigned, and executed locally, at edge or cloud nodes, or collaboratively across multiple agents. Second, task-aware information exchange determines what information should be transmitted, when communication should occur, and at what representation or compression level. Third, network resource management adapts spectrum, bandwidth, power, access opportunities, scheduling priorities, routing, and reliability levels according to task requirements rather than link-level metrics alone. Finally, multi-agent adaptation and coordination enables distributed embodied agents to coordinate their task roles, communication behaviors, and resource usage under heterogeneous and time-varying conditions. These control dimensions are coupled through the common objective of maximizing task-level utility under limited system resources.

Formally, at each decision epoch $t$, the system observes a state $s_t$ that captures task progress, agent conditions, network states, computing availability, and resource budgets. Based on $s_t$, the policy $\pi$ selects an action $a_t \in \mathcal{A}(s_t)$, which may involve task scheduling, information exchange, resource allocation, or multi-agent coordination. The immediate reward $r(s_t,a_t)$ evaluates the resulting task utility while accounting for communication, computation, energy, latency, and other operational costs. Since practical MEAN deployments are further subject to hard requirements on reliability, safety, latency, and resource availability, the orchestration problem can be formulated as a Constrained Markov Decision Process (CMDP)\cite{altman2021constrained}:
\begin{equation}
\begin{aligned}
\max_{\pi} \quad &
\mathbb{E}_{\pi, s_0 \sim \rho}
\left[
\sum_{t=0}^{\infty} \gamma^t r(s_t,a_t)
\right] \\
\text{s.t.} \quad &
\mathbb{E}_{\pi, s_0 \sim \rho}
\left[
\sum_{t=0}^{\infty} \gamma^t g_i(s_t,a_t)
\right]
\leq G_i, \quad \forall i \in \mathcal{I}, \\
& a_t \in \mathcal{A}(s_t), \quad
s_{t+1} \sim P(\cdot \mid s_t,a_t),
\end{aligned}
\label{eq:cmdp}
\end{equation}
where $\pi$ denotes the orchestration policy, $\gamma \in (0,1)$ is the discount factor, $P(\cdot|s_t,a_t)$ characterizes the dynamic evolution of the MEAN environment, and $g_i(s_t,a_t)$ represents the $i$-th constraint cost and $G_i$ denotes its admissible cumulative bound, e.g., for latency, energy consumption, interference, or safety risk.

The above formulation can be instantiated according to different levels of observability and control granularity. When a centralized controller has access to sufficient global state information, MEAN orchestration can be modeled as a standard Markov decision process. When individual agents observe only local sensing, communication, and task information, a partially observable MDP (POMDP) provides a more realistic formulation. Hierarchical RL can further separate high-level task orchestration from lower-level communication and resource control, thereby accommodating the different timescales of mission planning and network adaptation. For distributed MEAN deployments involving multiple cooperative or competing embodied agents, multi-agent reinforcement learning (MARL) provides a natural mechanism for decentralized decision-making and coordination\cite{hernandez2019survey}.

From this perspective, RL acts as a general adaptation and orchestration engine rather than an isolated wireless optimization tool. It links high-level embodied-task requirements with information exchange, network resources, computing capabilities, and multi-agent interactions, allowing MEAN to dynamically adjust its operation according to task progress and environmental feedback. Following this unified framework, the remainder of this section discusses three primary orchestration dimensions: RL-based task orchestration, task-aware information exchange, and task-aware network resource management. Advanced RL mechanisms are then introduced to support distributed coordination, rapid adaptation, safety-aware operation, and semantic-aware decision-making in dynamic MEAN deployments.

\subsection{RL for Task Orchestration}

Embodied AI tasks in MEAN are typically composed of multiple interdependent subtasks, such as perception, localization, mapping, inference, planning, control, and inter-agent collaboration. Because these stages impose different execution and resource requirements, task orchestration must determine how they are scheduled and where they should be executed across available agents and computing nodes\cite{park2019wireless,tang2020deep}. In addition, many subtasks depend on intermediate outputs from earlier stages, making their logical dependencies an important part of the orchestration process.

\begin{figure*}[!t]
    \centering
    \includegraphics[width=0.95\textwidth]{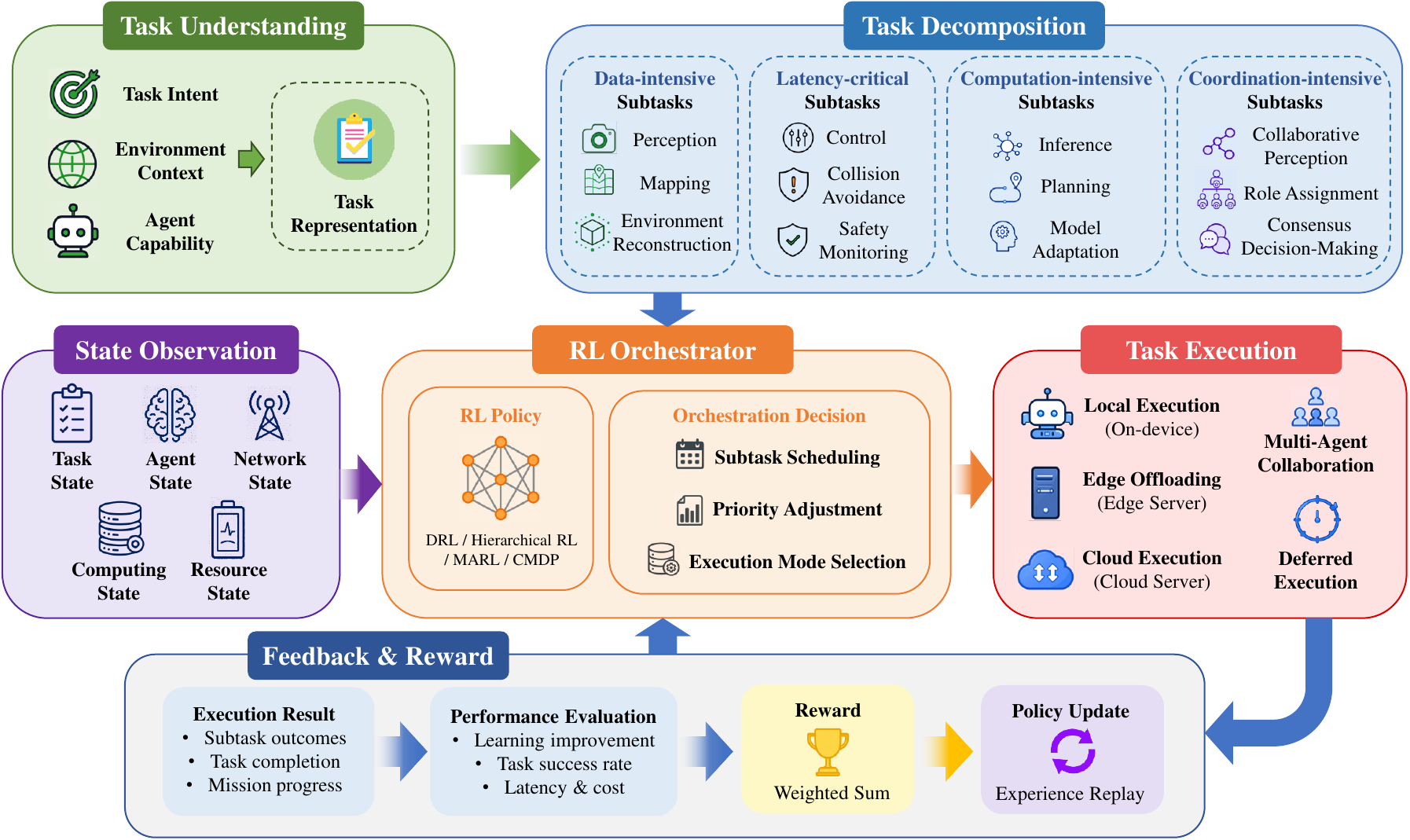}
    \caption{RL-enabled task orchestration framework for MEAN. Task understanding transforms task intent, environmental context, and agent capability into a structured representation, which is decomposed into heterogeneous subtasks. The RL orchestrator then maps observed task, agent, network, computing, and resource states to scheduling, priority, and execution-mode decisions, while execution feedback drives closed-loop policy refinement.}
    \label{fig:rl_task_orchestration}
\end{figure*}

RL provides a flexible mechanism for learning such orchestration policies from system states and execution feedback. As illustrated in Fig.~\ref{fig:rl_task_orchestration}, the orchestrator combines task structure with the current system state to determine how each subtask should be scheduled and executed. For example, latency-sensitive control may favor local execution, computation-intensive inference may be offloaded to an edge server, while collaborative perception may involve multiple agents. The policy can therefore adapt task execution to the current task and system conditions rather than relying on a fixed workflow.

Several RL formulations are applicable to task orchestration in MEAN:
\begin{itemize}
    \item \textbf{Hierarchical RL}: separates high-level task decomposition and execution planning from lower-level actions such as offloading, transmission scheduling, model selection, and sensing control\cite{dayan1992feudal}.
    \item \textbf{Constrained MDP}: incorporates explicit latency, energy, bandwidth, reliability, computation, or safety constraints into orchestration decisions\cite{zhao2023state}.
    \item \textbf{POMDP}: addresses incomplete or delayed observations of task, network, or agent states\cite{kaelbling1998planning}.
    \item \textbf{MARL}: supports distributed orchestration when multiple embodied agents make decisions based on local observations while jointly executing a global task\cite{oroojlooy2023review}.
\end{itemize}

In this setting, the RL state captures task progress together with the resources relevant to execution, while the action space specifies how subtasks are scheduled, prioritized, and assigned. The reward evaluates task progress against the resulting execution and resource costs. This allows the policy to coordinate both execution placement and task progression, particularly when later subtasks depend on timely outputs from earlier stages.

One representative direction is knowledge-graph-assisted RL for task decomposition. A knowledge graph can explicitly represent the structure and constraints of an embodied task, while RL uses this representation together with the system state to determine its decomposition\cite{zheng2024knowledge}. The resulting dependency graph provides a structured basis for subsequent scheduling and resource allocation. By making these dependencies explicit, the policy can reason over the task structure rather than treating subtasks as independent scheduling units. Such graph-based representations are particularly useful for embodied tasks with semantic or temporal dependencies that are difficult to capture using flat action spaces\cite{tenorth2013knowrob}.

Beyond task decomposition, the same orchestration framework can support execution-mode and priority decisions across heterogeneous agents and computing nodes. For example, computation-intensive subtasks may be assigned to local, edge, cloud, or collaborative execution according to available resources\cite{fan2024hybrid}, while MARL can coordinate role assignment and distributed execution among multiple agents\cite{zhu2024survey}.

Overall, RL-based task orchestration provides a mechanism for translating structured embodied tasks into adaptive scheduling, priority, and execution decisions, forming the task-level control layer of the broader RL-enabled MEAN framework.

\begin{figure*}[!t]
    \centering
    \includegraphics[width=0.9\textwidth]{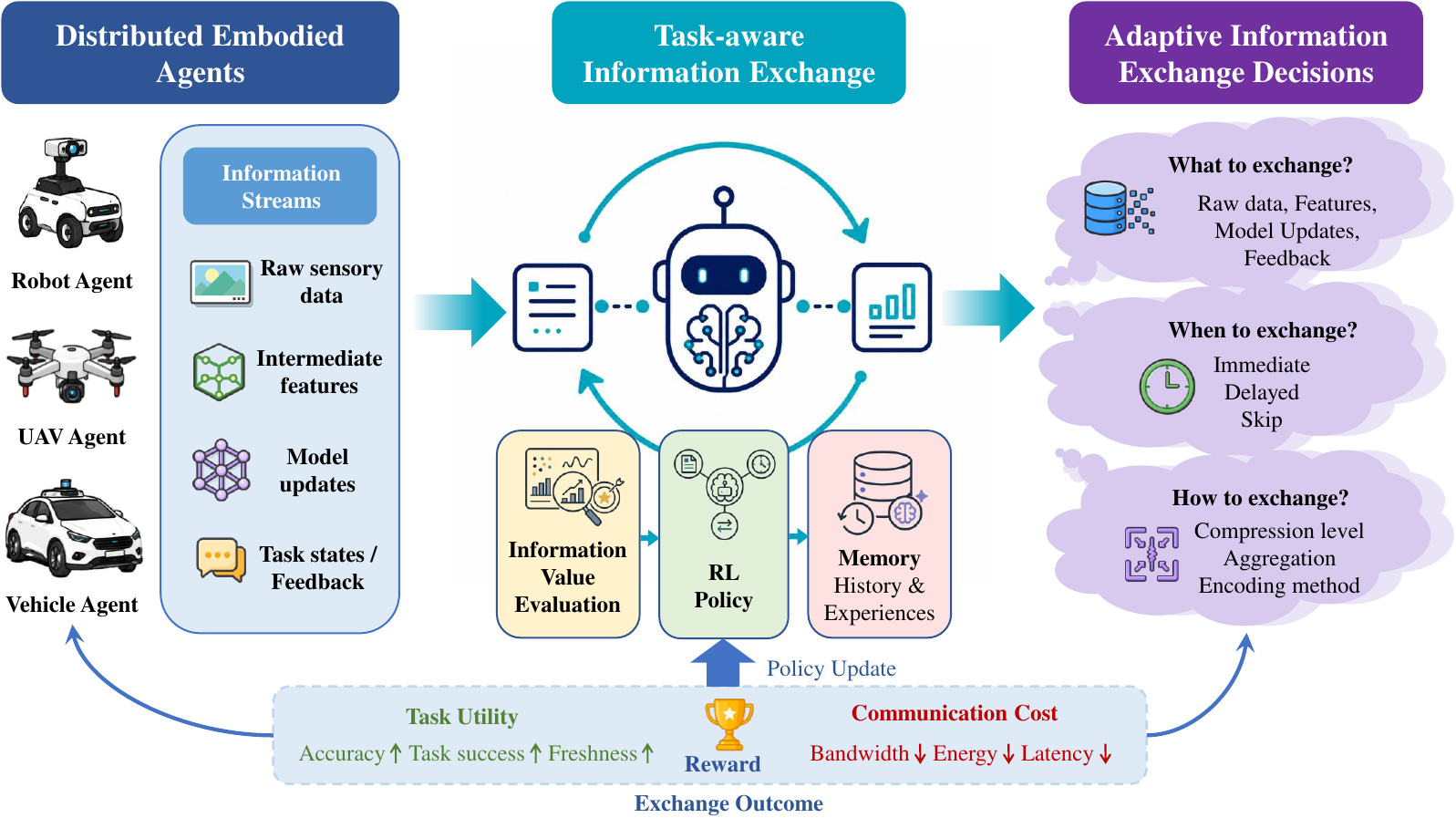}
    \caption{RL-enabled task-aware information exchange framework for MEAN. The RL policy evaluates information value, task state, and communication conditions to determine what information to exchange, when to transmit it, and how it should be represented. Exchange outcomes provide feedback for subsequent policy adaptation.}
    \label{fig:rl_information_exchange}
\end{figure*}

\subsection{RL for Information Exchange}

Information exchange connects distributed embodied task execution with wireless communication in MEAN. Embodied agents may share raw observations, learned features, model updates, inference results, and coordination-related information. However, the usefulness of these transmissions is highly context-dependent: information that is valuable at one stage of a task may become outdated at another. Its value may also depend on what the receiving agent already knows and what information is needed for its current decision. Since different information types vary in freshness, task relevance, and communication cost, transmitting all available information is often inefficient. RL-based information exchange therefore focuses on deciding what information to exchange, when to transmit it, and how it should be represented or compressed according to task and system states\cite{mcmahan2017communication}.

Unlike conventional traffic scheduling, the value of a transmission in MEAN depends on its contribution to task execution rather than data volume alone. For example, a compact but urgent control message may be more valuable than a large sensory stream, while intermediate features may replace raw observations when comparable inference performance can be maintained with lower communication cost\cite{liu2020when2com,kang2017neurosurgeon}. This makes information selection closely related to the downstream use of the transmitted content, rather than only to its size or channel requirement.

Two representative information-exchange patterns are particularly relevant. In model-distribution architectures, including split learning, collaborative inference, and model partitioning, the policy can adapt where computation is divided and what intermediate information is exchanged according to communication and computing conditions\cite{samikwa2022ares,lin2024efficient,li2024adaptive}. In distributed learning, such as federated and edge learning, RL can select participating agents and useful local updates based on their expected contribution and the resources required for communication\cite{nishio2019client,zhang2022multi}. This is especially relevant in MEAN, where observations collected by embodied agents are often heterogeneous and non-IID. As a result, communication decisions should account not only for whether information is available, but also for whether it provides new or complementary knowledge to the receiver.

As illustrated in Fig.~\ref{fig:rl_information_exchange}, the RL policy can jointly observe the progress of the task or learning process and the current communication and resource conditions. Based on the observed state, the policy decides whether communication is necessary and, if so, how it should be carried out, such as by adjusting update timing, compression, participation, peer selection, or model placement. The reward reflects the resulting task utility relative to the communication and execution cost. Feedback from downstream learning or inference can further indicate whether the transmitted information was useful, providing a basis for subsequent exchange decisions.

Different RL formulations can be adopted according to the information-exchange setting. Constrained RL can enforce communication, latency, energy, or reliability requirements; POMDP formulations can address incomplete observations of data and network states; MARL can coordinate distributed exchange among multiple agents; and actor--critic methods are suitable for continuous control variables such as compression ratio, transmission power, or update frequency.

\begin{figure*}[!t]
    \centering
    \includegraphics[width=0.95\textwidth]{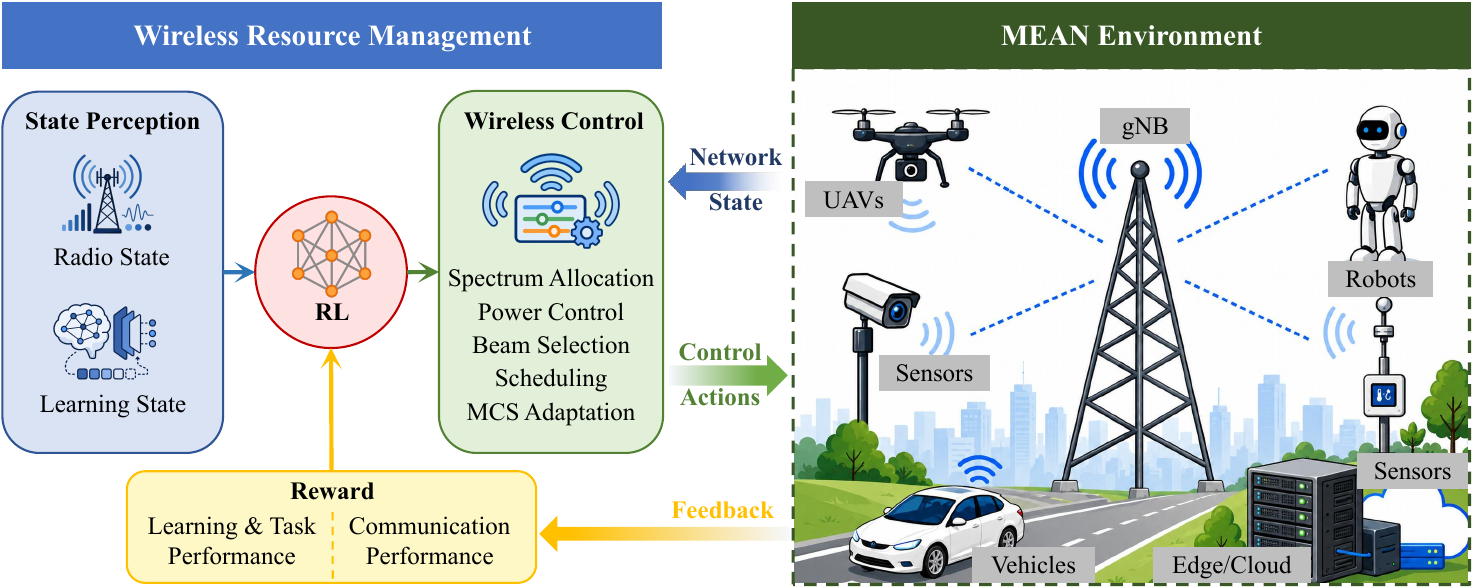}
    \caption{RL-enabled task-aware network resource management framework for MEAN. The RL controller observes network and task-related states and adapts radio resource allocation according to task and communication feedback.}
    \label{fig:rl_wireless_resource_management}
\end{figure*}

Adaptive update scheduling is a useful application of RL in distributed learning. Rather than transmitting model updates at fixed intervals, the policy can evaluate their current usefulness and decide whether communication is necessary or should be deferred\cite{zhang2023joint}. This helps reduce unnecessary synchronization while preserving valuable learning information.

RL can also support feature and feedback selection in collaborative inference. Depending on inference uncertainty, task relevance, and available bandwidth, agents may selectively transmit raw observations, intermediate features, semantic summaries, or confidence information\cite{chen2024deep}. Similarly, urgent feedback can be prioritized while redundant or low-value messages are suppressed or compressed.

Overall, RL-based information exchange enables MEAN to allocate communication resources according to the value of information to task execution and distributed intelligence, providing a task-aware interface between embodied agents and the underlying communication system.

\subsection{RL for Wireless Resource Management}

Wireless resource management in MEAN extends conventional radio optimization by incorporating task-level information into resource-control decisions. Rather than relying only on channel and traffic conditions, the controller also considers how communication affects task progress and the timeliness of learning or inference. The same amount of wireless resource may therefore have very different task value depending on which information is being served and when it is needed. Radio resources can thus be adapted according to their contribution to task execution instead of link-level performance alone.

This distinction becomes especially important when multiple embodied agents compete for limited wireless resources. A scheduler may need to balance the communication needs of agents that differ not only in channel quality, but also in their current role in the task. Consequently, resource allocation is no longer a simple matter of serving the user with the best instantaneous link condition. The controller must instead determine how limited radio resources should be shared so that communication supports the overall task process without unnecessarily degrading network efficiency.

As illustrated in Fig.~\ref{fig:rl_wireless_resource_management}, the RL controller jointly observes the current radio environment and task-related learning state. Based on these observations, it determines how communication resources should be allocated and configured across users and links. Such decisions affect not only transmission performance but also whether task-relevant information can reach the required agent or computing node in time. The reward therefore reflects the resulting task benefit together with communication efficiency and operational cost, while practical requirements on reliability, latency, interference, and resource availability can be incorporated as constraints. A key issue is how to relate radio-level outcomes to task-level performance. The effect of a resource decision may not be immediately visible at the communication layer, but may appear later through improved inference, fresher task information, or more timely control feedback. Designing suitable intermediate feedback is therefore important for allowing the RL controller to identify which radio decisions actually contribute to task progress.

In practice, these decisions may involve several radio-control variables that interact with one another. For example, changing the scheduling priority of a task flow may alter the bandwidth available to other users, while power or MCS adjustment can affect both transmission reliability and interference. RL is therefore useful for learning coordinated control policies over such coupled decisions instead of tuning each parameter independently. The control timescale can also vary: fast radio adaptation may respond to channel fluctuations, whereas task-aware priorities may change more slowly as the embodied task progresses.

Different RL methods can be selected according to the control space. Value-based methods such as DQN and DDQN are suitable for discrete decisions including scheduling, access, channel, or MCS selection\cite{ye2019deep,van2016deep}, whereas actor--critic methods such as DDPG and SAC are more appropriate for continuous variables such as power and bandwidth allocation\cite{haarnoja2018soft}. Constrained or safe RL can be used when strict latency, reliability, or safety requirements must be satisfied, while MARL supports distributed resource allocation among multiple embodied agents\cite{naderializadeh2021resource}. In practice, the choice of RL method depends not only on whether the control variable is discrete or continuous, but also on whether resource decisions must be coordinated across multiple users and task flows.

The deployment architecture also influences how resource-control policies are implemented. A centralized controller can exploit broader network and task information, while distributed policies may be more suitable when embodied agents must react locally or when global state exchange is costly. This trade-off makes MARL particularly relevant to large-scale MEAN deployments. MARL-based joint adaptive modulation and coding (AMC) and resource allocation provides a useful example of how distributed learning can support wireless control, where users learn resource-allocation policies from local observations and system feedback\cite{nasir2019multi}. While the original formulation mainly targets throughput and fairness, its objective can be extended in MEAN to account for the contribution of radio-resource decisions to task execution. In this case, users carrying more time-sensitive or task-critical information can be treated differently even under similar radio conditions, providing a natural transition from link-centric to task-aware control.

Overall, RL-based wireless resource management provides the network-level control layer of MEAN, allowing conventional radio resources to be dynamically allocated according to both communication conditions and task requirements.

\subsection{Advanced RL for Dynamic and Distributed MEAN}

Beyond task orchestration, information exchange, and wireless resource management, practical MEAN systems must also cope with distributed decision-making, changing operating conditions, and safety-critical execution. These challenges motivate more advanced RL mechanisms that extend the basic orchestration framework.

A first challenge arises from the distributed nature of MEAN. Multiple embodied agents often make decisions based on local observations while sharing communication and computing resources. MARL is therefore particularly relevant for coordinating their task execution, information exchange, and resource use\cite{lowe2017multi}. Centralized training with decentralized execution (CTDE) offers a practical implementation paradigm, although communication overhead, non-stationarity, and credit-assignment issues may still limit scalability.

Distributed coordination alone is insufficient when the operating environment changes over time. Variations in tasks, mobility, channel conditions, or resource availability can reduce the effectiveness of policies trained under fixed settings. Meta-RL can improve adaptation to new conditions with limited interaction data\cite{finn2017model}, while digital twins provide virtual environments for policy training, evaluation, and refinement before or during deployment\cite{khan2022digitaltwin}. Together, they can reduce the cost and risk associated with online adaptation.

As policies adapt to changing conditions, safety becomes another critical requirement. Safe and constrained RL is important when decisions must satisfy strict requirements on reliability, latency, energy, or physical safety. It helps the policy avoid unsafe actions and keep system operation within acceptable limits. This is especially important when RL is used for physical control or other safety-critical tasks.

Beyond coordination, adaptation, and safety, MEAN must also determine which information is worth communicating. Semantic-aware RL introduces task relevance and information value into communication decisions, allowing agents to adapt what information is exchanged and at what level of representation. This provides a natural bridge from RL-based communication control to the token-oriented communication framework introduced in the next section.

Together, these advanced RL mechanisms extend the basic MEAN orchestration framework from single-policy adaptation toward distributed, adaptive, safe, and semantically informed operation.

\section{Task-Oriented MEAN Based on Token Communication}
\begin{figure*}[t]
\centering
\includegraphics[width=0.95\textwidth]{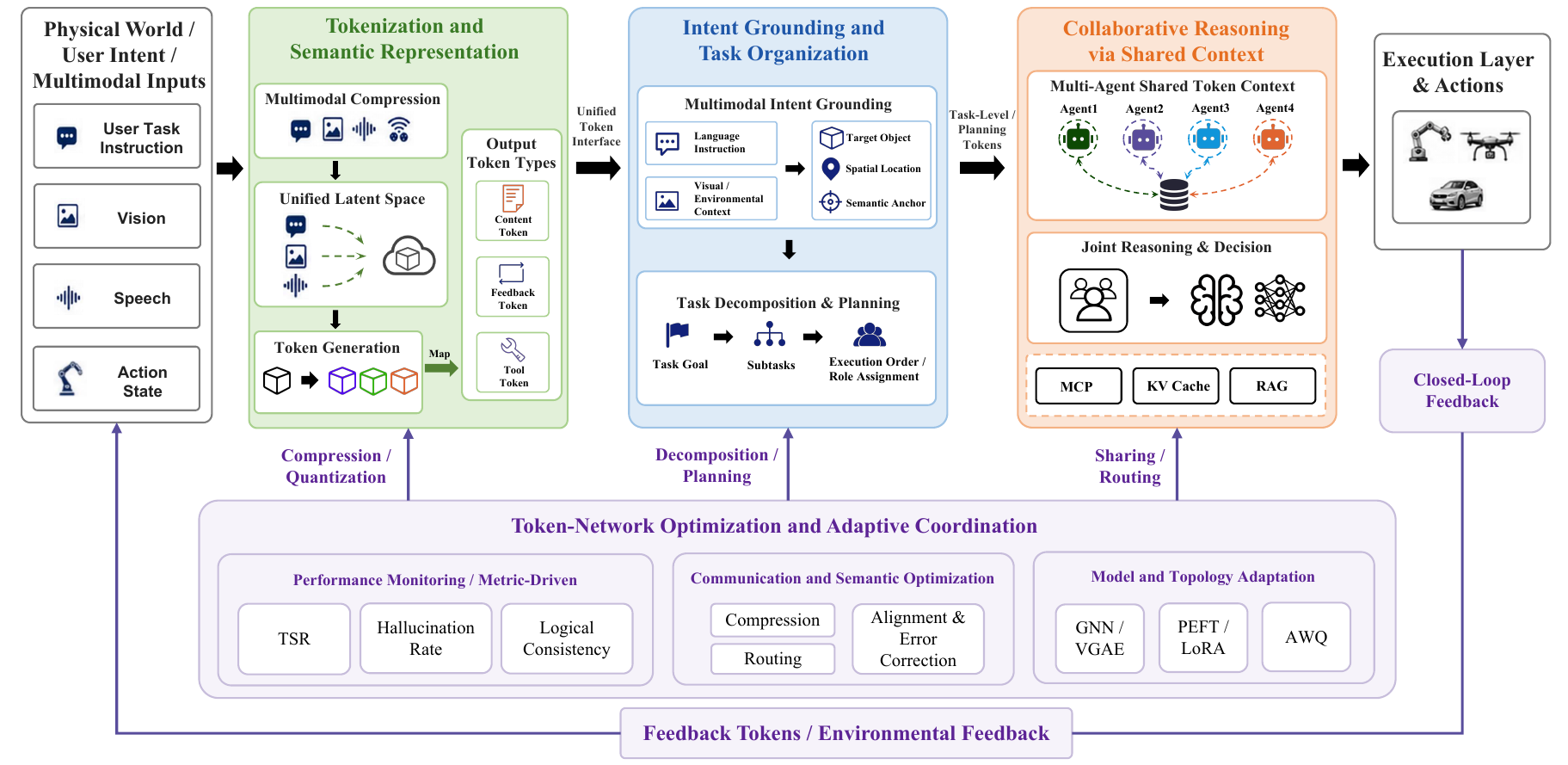}
\caption{Task-oriented token communication in MEAN. Multimodal observations, task intent, and feedback are encoded as semantic tokens for intent grounding, decomposition, and planning. Agents exchange shared token contexts for collaborative reasoning and coordinated action, supported by MCP-based interaction, KV-cache reuse, and RAG. Task outcomes feed back into token-network optimization, which adapts semantic exchange, routing, and model deployment.}
\label{fig:task-oriented_mean}
\end{figure*}

Section \ref{RL-enabled MEAN} examined how RL enables adaptive task orchestration, information exchange, and resource allocation in MEAN. Complementing this decision-making perspective, this section focuses on how task-relevant information is represented, aligned, and exchanged through tokens to support collaborative reasoning and coordinated task execution.

Task-oriented token communication shifts the emphasis from bit-level reconstruction to the effective exchange of information needed for task completion. By encoding multimodal observations, task intent, and execution feedback into compact semantic representations, tokens support intent understanding, task decomposition, and collaboration among distributed embodied agents. As illustrated in Fig.~\ref{fig:task-oriented_mean}, this architecture connects perception, reasoning, and action through token exchange, supporting coordinated task execution under communication and computing constraints.

\subsection{From Bit-Stream to Token Communication: A Paradigm Shift}
Reliable bit delivery remains essential, but MEAN also requires efficient information exchange under high-volume multi-modal sensing, limited edge bandwidth, and stringent action deadlines. Transmitting complete observations or relying on centralized processing may delay distributed decisions, while heterogeneous agents need interoperable representations. Token communication addresses these constraints by selecting information according to its relevance to downstream reasoning and physical execution, rather than treating all source details as equally important.

\subsubsection{Definition and Connotation of Tokens}
\begin{figure}
    \centering
    \includegraphics[width=1\linewidth]{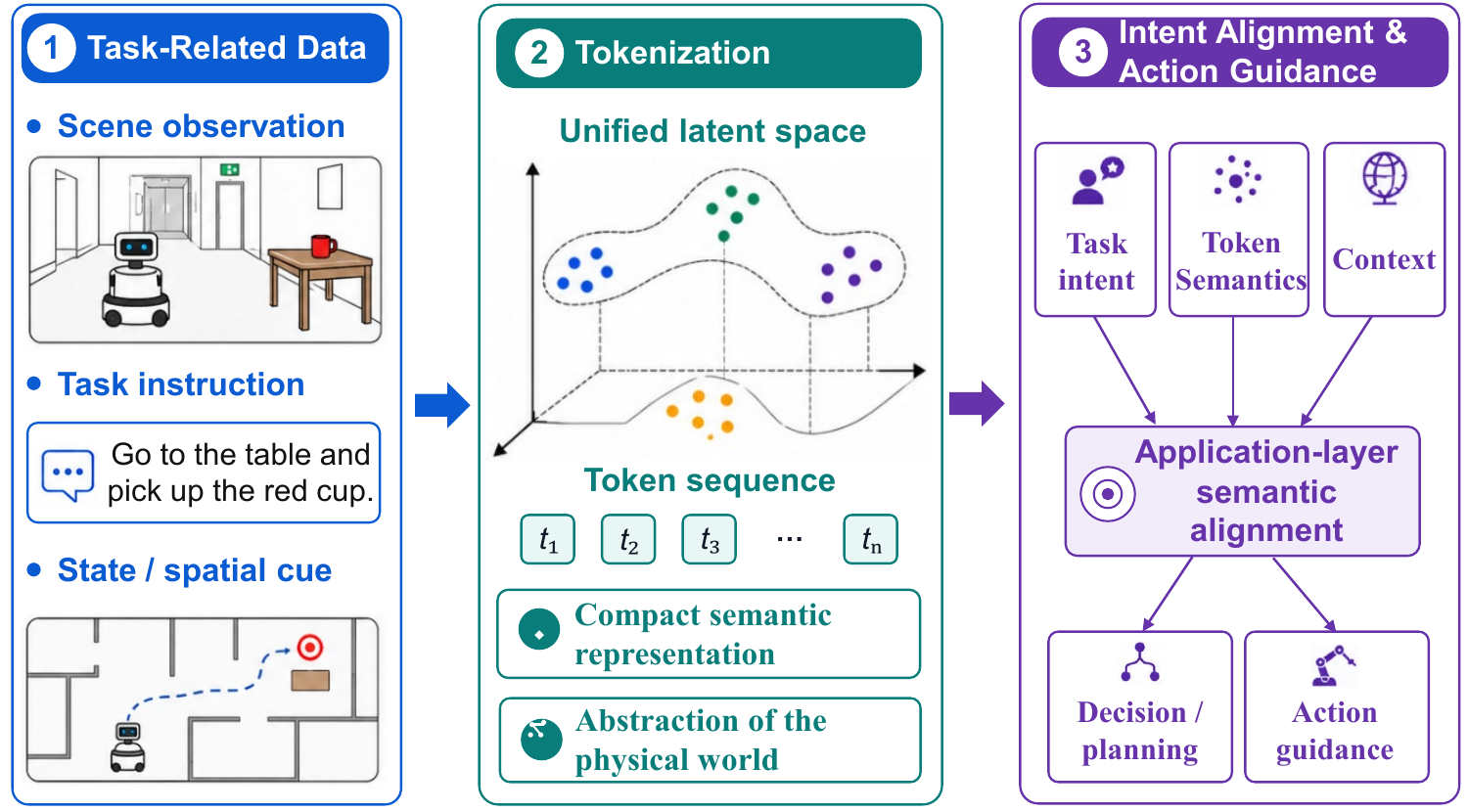}
    \caption{Token definition and utilization in embodied intelligence. Task-related observations are encoded into compact semantic tokens, which are aligned with task intent and environmental context to support planning and action.}
    \label{fig:part_four_token_definition}
\end{figure}

Within the MEAN framework, a token is no longer merely a basic text unit processed by large language models (LLMs), but a semantic abstraction of the physical world perceived by embodied agents as shwon in Fig.~\ref{fig:part_four_token_definition}. This redefinition enables tokens to transcend simple symbolic attributes, becoming a core nexus connecting the physical perception space with the cognitive planning space. Essentially, a token serves as a universal cognitive carrier for embodied agents within the perception-reasoning-action loop. It encompasses compressed representations of multi-modal features mapped through a unified latent space, while carrying specific intent instructions and task constraints. Through the transition from primitive bitstream transmissions to structured semantic tokenization, the representation and flow of information within embodied networks are fundamentally reshaped, imbuing the communication process with a high cognitive dimension. This supports a transition in end-to-end collaborative logic from physical-layer signal replication to application-layer intent alignment.

\subsubsection{Multi-modal Feature Compression}
Multi-modal feature compression is essential for exchanging task-relevant information under limited bandwidth. Rather than transmitting full sensory streams, MEAN maps high-dimensional observations into compact representations in a shared latent space. Three mechanisms support this task-oriented compression:

\begin{itemize}
\item \textbf{Semantic Bottleneck Constraints}: Heavily inspired by the Information Bottleneck (IB) theory, the core logic of this mechanism lies in trading off irrelevant raw details for high compression ratios and generalization capabilities. Unlike traditional source coding governed by the Mean Squared Error (MSE) criterion, the tokenization process in MEAN follows a task-driven informational utility objective. The Perception Module acts as a non-linear encoder, mapping high-dimensional data—such as LiDAR point clouds, visual imagery, and RF signals—onto a low-dimensional task-optimal manifold. Under these constraints, the agent extracts only the specific semantic features essential for decision-making, such as object categories, relative poses, and dynamic trajectories, while suppressing task-irrelevant background artifacts and stochastic environmental noise. This reduces bandwidth requirements by orders of magnitude and enhances system robustness, ensuring the precise delivery of primary intents under extreme resource constraints \cite{tishby2000informationbottleneckmethod, zhang2025agenticaigenerativeinformation}.

\item \textbf{Cross-modal Alignment and Representation}: To facilitate inter-agent semantic coherence among agents equipped with heterogeneous sensors, MEAN utilizes multi-modal large language models (VLMs) as the underlying architecture to construct a Unified Latent Space. This mechanism employs techniques such as contrastive learning to encode information from diverse modalities, including vision, haptics, and array signals, into a consistent token representation system. This aligned representation achieves latent-space semantic invariance: regardless of whether the input originates from optical sensors or radar depth maps, the generated tokens maintain topological proximity within the shared latent manifold. Such high semantic faithfulness is critical for handling Partially Observable Markov Decision Processes (POMDP). In extreme environments with sensory occlusion, cross-modal alignment enables the system to utilize tokens from alternative modalities, such as radio frequency (RF) signals or light detection and ranging (LiDAR), to autonomously perform semantic context completion, maintaining perceptual continuity and ensuring task execution success in complex dynamic scenarios \cite{radford2021learningtransferablevisualmodels, 2023GPT4VisionSC}. 

\item \textbf{Adaptive Quantization and Pruning}: To adapt to resource-constrained edge hardware, MEAN employs activation-aware weight quantization (AWQ) \cite{lin2024awqactivationawareweightquantization} and structural pruning techniques. By observing activation distributions during inference, the system identifies performance-critical weight subsets—those that occupy a minimal fraction of the total parameters yet exert a dominant influence on inference accuracy. High-precision feature tokens are then compressed into low-bitwidth representations, such as INT4, optimizing transmission efficiency while maintaining functional fidelity with minimal accuracy degradation.
\end{itemize}

\subsubsection{Driving Force of Reasoning and Decision Making}
In the hierarchical logical architecture of MEAN, tokens serve as core computational primitives that span perceptual boundaries and action spaces. Acting as a hub connecting physical reality with cognitive logic, tokens trigger the dynamic evolution of the embodied "Perception-Reasoning-Action" loop through the semantic instantiation of multi-modal inputs. This token-based logic drive enables agents to transform fragmented environmental perceptions into a continuous stream of task planning, thereby maintaining decision consistency in dynamic environments.

\begin{itemize}
\item \textbf{Task-Token Driven Planning}: Upon receiving an intent token, the cognitive core utilizes chain-of-thought (CoT) reasoning to decompose it into a sequence of task sub-goal tokens. Each task token carries specific logical instructions, driving the agent to perform long-horizon path planning and tool invocation.
\item \textbf{Closed-loop Feedback and Adaptive Strategies}: The decision stream formed by tokens can seamlessly integrate feedback tokens from the environment. For instance, the physical state of the execution end, including parameters such as gripping torque or movement position, is transformed in real-time into feedback tokens and sent back to the cognitive layer. Through a self-refinement cycle, the reasoning direction is dynamically corrected.
\item \textbf{Tool Invocation and External API Triggering}: When local cognitive capabilities are insufficient for specialized tasks, MEAN introduces an automated tool invocation mechanism. The Toolformer mechanism, through self-supervised learning, enables the model to autonomously insert specific API call tokens into the generated token stream, triggering on-demand calls to external utilities such as calculators, search engines, or specialized sensors \cite{schick2023toolformerlanguagemodelsteach}. Furthermore, when integrated with the Gorilla framework, MEAN nodes can autonomously retrieve and invoke an extensive array of external RESTful APIs, effectively mitigating the inherent constraints of LLMs regarding domain-specific knowledge and numerical precision \cite{patil2023gorillalargelanguagemodel}. This augmented reasoning paradigm enables embodied agents to leverage the Reasoning and Acting (ReAct) framework to facilitate iterative action-observation cycles and real-time decision refinement based on the functional tokens retrieved from external computational interfaces \cite{yao2023reactsynergizingreasoningacting}.
\end{itemize}

\subsubsection{Transformation of Communication Paradigms}
The core logic of the MEAN communication protocol is undergoing a fundamental transformation from transmitting how to execute (bits) to what to execute (intent/tokens). This shift marks a paradigm leap from the traditional internet of things (IoT), where devices are merely responsible for perception and transmission, to the internet of agents (IoA), where edge nodes possess capabilities for perception, reasoning, planning, and independent action \cite{Wang_2026}.

\begin{itemize}
\item \textbf{Intent-driven Interaction}: In contrast to the bit-oriented paradigm focused on precise replication of waveforms or bit sequences, MEAN prioritizes intent understanding. Embodied agents exchange high-level semantic tokens, allowing the receiver to autonomously infer subsequent actions based on local semantic cues. This significantly reduces the dependency on continuous, high-intensity raw data exchange, thereby drastically decreasing communication overhead \cite{zhang2025agenticaigenerativeinformation, liu2025lametaintentawareagenticnetwork}.
\item \textbf{Cognitive Consistency under Resource Constraints}: In 6G edge environments, achieving full physical-layer synchronization and global state visibility is difficult due to channel jitter, limited bandwidth, and dynamic network topologies. To address this, MEAN introduces the model context protocol (MCP) \cite{krishnan2025advancingmultiagentsystemsmodel}. Distributed embodied agents share key task tokens and context representations via MCP to ensure consistent contextual understanding across different nodes. This ensures that group collaboration, such as coordinated coverage by UAV swarms or dynamic scheduling in vehicular networks, maintains high goal-driven coherence and task execution robustness under low-bandwidth, non-stationary conditions \cite{10811956, tong2024wirelessagentlargelanguagemodel}.
\end{itemize}

\subsection{Intent Grounding and High-Level Task Decomposition}
Intent grounding connects language-level task goals with physical entities, spatial context, and agent capabilities. In MEAN, this requires aligning instructions with multi-modal observations and execution constraints so that the resulting task tokens can support feasible decomposition and planning in dynamic, partially observable environments.

\subsubsection{Multi-modal Intent Grounding}
Within the MEAN interaction ecosystem, intent grounding transcends unimodal linguistic extraction; instead, it manifests as a deep tensor fusion and feature alignment between multi-modal perceptual streams and the cognitive core. The system synchronizes captured visual scenes, acoustic commands, and RF perceptual data in the spatio-temporal domain for integration into the embodied cognitive core. This fusion mechanism endows the system with context-aware situational comprehension, enabling the disambiguation of underspecified commands based on the real-time physical environment.

\begin{itemize}
\item \textbf{Context-aware Intent Instantiation}: The Perception Module provides the requisite physical references for semantic instructions through the cross-verification of real-time visual features, acoustic signals, and RF fingerprints. For instance, upon receiving an abstract command regarding the traversal of a designated region, the system utilizes spatial tokens to anchor the intent to a specific local coordinate frame and its associated physical parameters. Global correlation between linguistic query vectors and visual feature maps is executed via cross-attention mechanisms. By constructing a multi-modal alignment score matrix, the system identifies the physical properties and spatial poses of specific objects referenced in the command within the current local coordinate system. This process typically leverages frameworks such as Grounded Language-Image Pre-training (GLIP) or Grounding DINO, unifying visual object detection and semantic comprehension under a singular objective function to facilitate the real-time anchoring of intent within the physical domain \cite{li2022groundedlanguageimagepretraining, liu2024groundingdinomarryingdino}.
\item \textbf{Task Tokenization}: The cognitive core maps environment-augmented instructions into a sequence of structured task tokens. Each token represents a logical atom and implicitly includes the boundary conditions and physical constraints of the action. MEAN transforms intent into executable code blocks or policy tokens, allowing agents to leverage the reasoning power of pre-trained models to generate logically rigorous low-level instructions.
\end{itemize}

\subsubsection{Task Decomposition and Chain-of-Thought Planning}
Faced with long-horizon tasks characterized by high non-stationarity and complex physical constraints, MEAN must possess the ability to deconstruct macro-intents into micro-action primitives. This process primarily relies on hierarchical task topology construction and a token-sequence-based CoT reasoning mechanism.

\begin{figure*}
    \centering
    \includegraphics[width=0.95\linewidth]{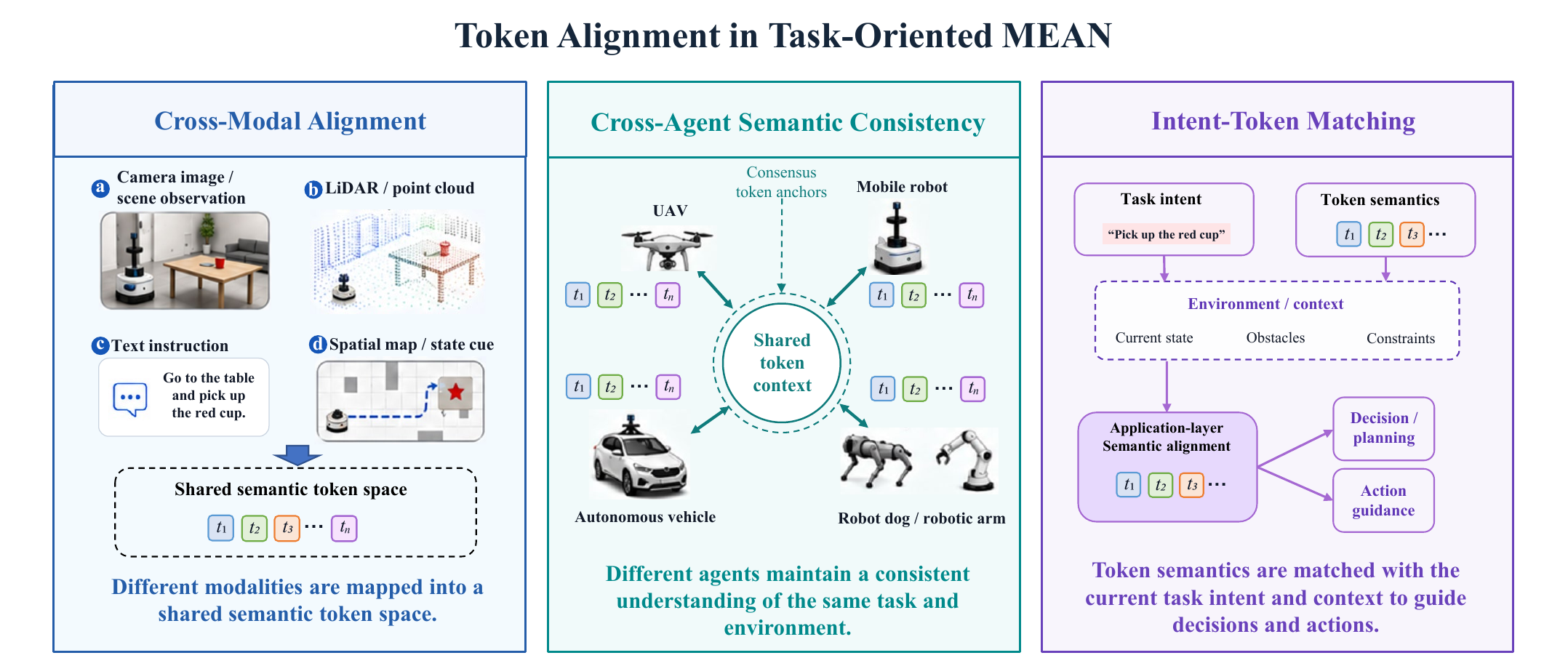}
    \caption{Three aspects of token alignment in task-oriented MEAN, including cross-modal alignment in a shared token space, cross-agent semantic consistency, and intent-token matching for decision-making and action guidance.}
    \label{fig:part_four_token_alignment}
\end{figure*}

\begin{itemize}
\item \textbf{Hierarchical Task Topology Construction}: MEAN utilizes large-scale models as centralized decision-making engines to recursively decompose high-level abstract goals into a series of sub-goal tokens characterized by rigorous causal dependencies. Diverging from unimodal linguistic planning, MEAN incorporates a physical affordance assessment mechanism during the decomposition phase. When generating individual task tokens, the cognitive core evaluates not only the semantic coherence of the plan but also the operational feasibility of the action within the current physical state, integrating environmental tokens facilitated by the perception module \cite{ahn2022icanisay}. Consequently, the system constructs a dynamic task topology, ensuring that complex long-horizon planning is consistently grounded within the feasible operational envelope. This approach effectively mitigates cognitive hallucinations and semantic-physical inconsistencies that are prevalent when traditional large-scale models are deployed in embodied scenarios.
\item \textbf{Dual-constrained CoT Reasoning}: To facilitate the seamless mapping from discrete symbolic logic to continuous actuation, MEAN employs an enhanced CoT reasoning protocol. By integrating formal logical predicates and spatio-temporal constraint parameters into the token stream, the architecture implements an internalized self-reflective reasoning process for task execution \cite{huang2022innermonologueembodiedreasoning}. Prior to the dispatch of each task token to the actuation layer, the cognitive core performs closed-loop verification to ensure the task sequence preserves causal coherence within the temporal domain and adheres to kinematic and collision-avoidance constraints in the spatial dimension. This modular planning mechanism enables agents to adapt to stochastic environmental perturbations by selectively updating affected token nodes rather than necessitating global plan re-computation, thereby enhancing computational efficiency and operational robustness.
\item \textbf{Multi-agent Token Chaining}: In scenarios requiring multi-agent swarm intelligence, such as coordinated reconnaissance by UAV swarms, MEAN facilitates orchestrated task distribution and cross-node synchronization. The leader node defines the operational constraints and boundary conditions for the motion primitives of collaborative nodes through the generation of structured path-planning token clusters. By leveraging the end-to-end mapping capabilities of Robotic Transformer 2 (RT-2), these high-level semantic tokens are translated into low-level actuation tokens. This architecture enables heterogeneous agents to function as an integrated decentralized execution system, completing complex spatio-temporal collaborative tasks via a deterministic and sequential propagation of token-mediated control signals \cite{brohan2023rt2visionlanguageactionmodelstransfer, zhang2025agenticaigenerativeinformation}.
\end{itemize}

\subsection{Collaborative Reasoning via Token-based Context Sharing}
Distributed embodied agents must maintain compatible task contexts despite partial observations, intermittent links, and changing network topologies. Exchanging selected context and state tokens can reduce repeated synchronization of raw histories while supporting shared interpretation of task goals and constraints. This motivates token-based context sharing for collaborative reasoning and coordinated execution.

\subsubsection{Model Context Protocol (MCP)}
To maintain distributed cognitive consistency in edge scenarios with limited communication resources, MEAN introduces MCP. Acting as a standardized interaction framework for semantic alignment among distributed agents, the protocol achieves low-loss synchronization of cognitive states through the structural decoupling of reasoning contexts.

\begin{itemize}
\item \textbf{Incremental Update of Cognitive States}: Instead of retransmitting entire reasoning histories, agents may exchange updated task intents and context tokens. Prompt caching and compatible cross-node KV-cache reuse can further reduce redundant processing and communication when agents share suitable model and representation interfaces \cite{krishnan2025advancingmultiagentsystemsmodel}. The resulting savings depend on cache compatibility, update size, and synchronization requirements.
\item \textbf{Semantic Consistency Verification}: To address cognitive bias between heterogeneous embodied platforms, such as robots with different perceptual modalities or computational gradients, the MCP protocol establishes a unified semantic space baseline. By introducing Consensus Token Anchors, the system can formally verify key task predicates, including spatial constraints and action boundaries. This mechanism ensures that even with varying model parameter scales, nodes can reach a logical consensus on the same task intent, fundamentally reducing the probability of conflict in physical collaboration \cite{liu2025lametaintentawareagenticnetwork}.
\end{itemize}

\subsubsection{Distributed Token Retrieval (RAG)}
To address long-term dependencies and non-stationary topology challenges in embodied tasks, MEAN integrates a distributed retrieval-augmented generation (RAG) mechanism, constructing a distributed memory network with spatio-temporal correlation.

\begin{itemize}
\item \textbf{Context Token Encoding}: Embodied agents do not need to maintain redundant global trajectories; instead, they encode historical observations and decision logic into Context Tokens with semantic features. Using GraphRAG technology, these tokens are organized into knowledge graphs with topological correlations. During complex long-horizon tasks, agents achieve precise retrieval of relevant historical experiences and environmental common sense through similarity comparisons in high-dimensional vector spaces, significantly enhancing the generalization ability of the decision module when facing low-probability edge cases \cite{radford2021learningtransferablevisualmodels}.
\item \textbf{Topology Adaptation and Robust Reasoning}: In the context of frequent 6G edge topology reconfiguration, MEAN adopts an asynchronous context synthesis strategy. When a node re-enters the network after intermittent communication disruption, the system utilizes semantic reconstruction techniques to recall critical task state tokens from the vector caches of neighboring nodes. By introducing a self-reflective retrieval mechanism (Self-RAG), agents can perform faithfulness verification on retrieved tokens, ensuring the collaborative loop maintains logical rigor and operational stability under conditions of incomplete information observation \cite{zhang2025agenticaigenerativeinformation}.
\end{itemize}

\subsection{Optimization and Performance Metrics for Token Networks}
Token-network optimization must balance semantic fidelity, task performance, communication overhead, and heterogeneous edge-computing resources. This section considers compact model support for token processing and task-aware metrics for evaluating communication efficiency and semantic reliability.

\subsubsection{Compact Model Support and Edge Deployment}
Given the power consumption and memory bandwidth limitations of mobile embodied devices, deploying large-scale models with full parameters is often impractical. Therefore, utilizing compaction techniques to build small language models (SLMs) or compact embodied models is a critical technical path for ensuring that edge nodes can generate high-quality reasoning tokens in real-time.

\begin{itemize}
\item \textbf{Multi-dimensional Model Compression}: MEAN adopts a combined strategy of quantization and structural pruning. By introducing techniques such as AWQ, models can run at low bit-width precision, significantly reducing memory footprint while maintaining the faithfulness of semantic reasoning \cite{lin2024awqactivationawareweightquantization}. Additionally, by removing redundant attention heads or neuron layers, pruning techniques further reduce the Floating Point Operations (FLOPs) per token inference, ensuring that embodied nodes maintain millisecond-level intent response speeds even with low computational power \cite{ma2023llmprunerstructuralpruninglarge}.
\item \textbf{Parameter-Efficient Fine-Tuning (PEFT) and LoRA Adaptation}: To enable compact models to handle specific embodied tasks, such as navigation in specific terrains or industrial gripping, MEAN extensively applies parameter-efficient fine-tuning (PEFT) and low-rank adaptation (LoRA) technology. By training only a tiny fraction of low-rank decomposition matrices, LoRA allows nodes to quickly adapt to new physical constraints and task intents without altering the pre-trained model backbone. This approach reduces training energy consumption and allows agents to achieve rapid knowledge transfer by exchanging small adapter weights \cite{hu2021loralowrankadaptationlarge}.
\end{itemize}

\subsubsection{Key Performance Metrics}
Conventional throughput and packet-loss metrics do not capture whether exchanged tokens preserve task-relevant meaning. We therefore consider three complementary measures: task success rate (TSR), communication reduction ratio (CRR), and hallucination rate (HR), together reflecting task effectiveness, transmission efficiency, and semantic reliability.

\begin{itemize}
\item \textbf{Task Success Rate (TSR)}: As the primary metric for measuring MEAN communication efficiency, TSR is defined as the probability that an embodied agent eventually achieves its preset physical goal driven by a specific token sequence
\begin{equation}
\mathrm{TSR}
=
\frac{1}{N}
\sum_{i=1}^{N}
\mathbb{I}\!\left(\mathcal{T}_i=\mathrm{success}\right),
\end{equation}
where $N$ denotes the total number of task trials, 
$\mathcal{T}_i$ represents the outcome of the $i$-th embodied task, 
and $\mathbb{I}(\cdot)$ is the indicator function, which equals $1$ 
if the task is successfully completed and $0$ otherwise. Unlike bit accuracy, TSR focuses on the correctness of logical outcomes. As long as the key intent tokens are correctly received and trigger a compliant action sequence, the system maintains a high TSR even if there are losses of non-critical characters during transmission.
\item \textbf{Communication Reduction Ratio (CRR)}: CRR is used to quantify the spectrum resources saved by semantic token transmission compared to original multi-modal data, such as video streams or point clouds. In embodied collaborative scenarios, tokens are high-level abstractions of the environment, and their transmission data volume is typically several orders of magnitude lower than raw signals. A high CRR indicates that MEAN can maintain complex swarm collaborative decision-making even in environments with extremely poor channel conditions or narrow bandwidths \cite{zhang2025agenticaigenerativeinformation}.
\item \textbf{Hallucination Rate (HR) and Logical Consistency}: This metric evaluates the frequency of logical errors, physical constraint violations, or false intents generated during token transmission and reasoning. Since large models may produce hallucinations under constrained computation, real-time monitoring of the hallucination rate in the token stream is crucial for ensuring the safety of embodied tasks
\begin{equation}
\mathrm{HR}
=
\frac{1}{N}
\sum_{i=1}^{N}
\mathbb{I}
\left(
\hat{y}_i \not\models
\mathcal{K}_i \cup \mathcal{C}_i
\right),
\label{eq:hallucination_rate}
\end{equation}
where $N$ denotes the total number of evaluated outputs,
$\hat{y}_i$ is the output generated in the $i$-th task,
$\mathcal{K}_i$ denotes the available factual or environmental
knowledge, and $\mathcal{C}_i$ represents the task and physical
constraints. The indicator function $\mathbb{I}(\cdot)$ equals
$1$ when the generated output is unsupported by the available
knowledge or violates the specified constraints, and $0$
otherwise. By combining Self-RAG with physical simulation feedback, MEAN can quantify the alignment between the reasoning chain and physical laws, triggering necessary fault-tolerance or retransmission mechanisms at the execution end \cite{asai2023selfraglearningretrievegenerate, liu2025lametaintentawareagenticnetwork}.
\end{itemize}

\section{Task-driven Deployment of MEAN based on Large AI Model}

The fundamental motivation for deploying Large AI Model (LAM) on MEAN stems from the dual requirements of embodied intelligence tasks: real-time closed-loop control and high-dimensional semantic understanding. In traditional cloud-based inference paradigms, raw sensory data must traverse a complete uplink transmission path to be processed in the cloud, after which control commands are sent back down to the device. This process incurs end-to-end latency typically ranging from hundreds of milliseconds to several seconds, a delay that fails to meet the stringent real-time demands of mobile embodied agents performing tasks such as obstacle avoidance or grasping control. Concurrently, complex embodied manipulation scenarios require LAM to perform joint reasoning over multi-modal data (visual, linguistic, tactile, etc.), imposing computational loads that frequently exceed the capacity of a single chip on a mobile terminal. Consequently, MEAN must seek a novel deployment paradigm that balances model capability against device constraints, which avoids sacrificing real-time performance by placing LAM entirely in the cloud, while also circumventing the computational and power limitations of deploying them wholly on the edge \cite{zheng2026roboecc, han2026pipesd}.

However, deploying LAM in MEAN scenarios requires careful consideration of the following issues: the design of the deployment architecture, token-based communication mechanisms between nodes, the capability for real-time inference execution, and continuous deployment optimization in dynamic environments. Specifically, the deployment architecture determines the specific LAM deployment pattern in the cloud-edge-device system. Token-based communication mechanisms enable efficient information flow of tokens as the fundamental unit between different nodes to meet the requirements of real-time tasks. Real-time inference execution ensures that LAM execution satisfies the time budgets of real-world multi-agent task execution. Continuous deployment optimization allows the system to adaptively adjust as environments and resource states evolve.

\subsection{Cloud-Edge-Device and Multi-Agent Deployment Architectures}

\begin{figure*}[t]
    \centering
    \includegraphics[width=0.85\textwidth]{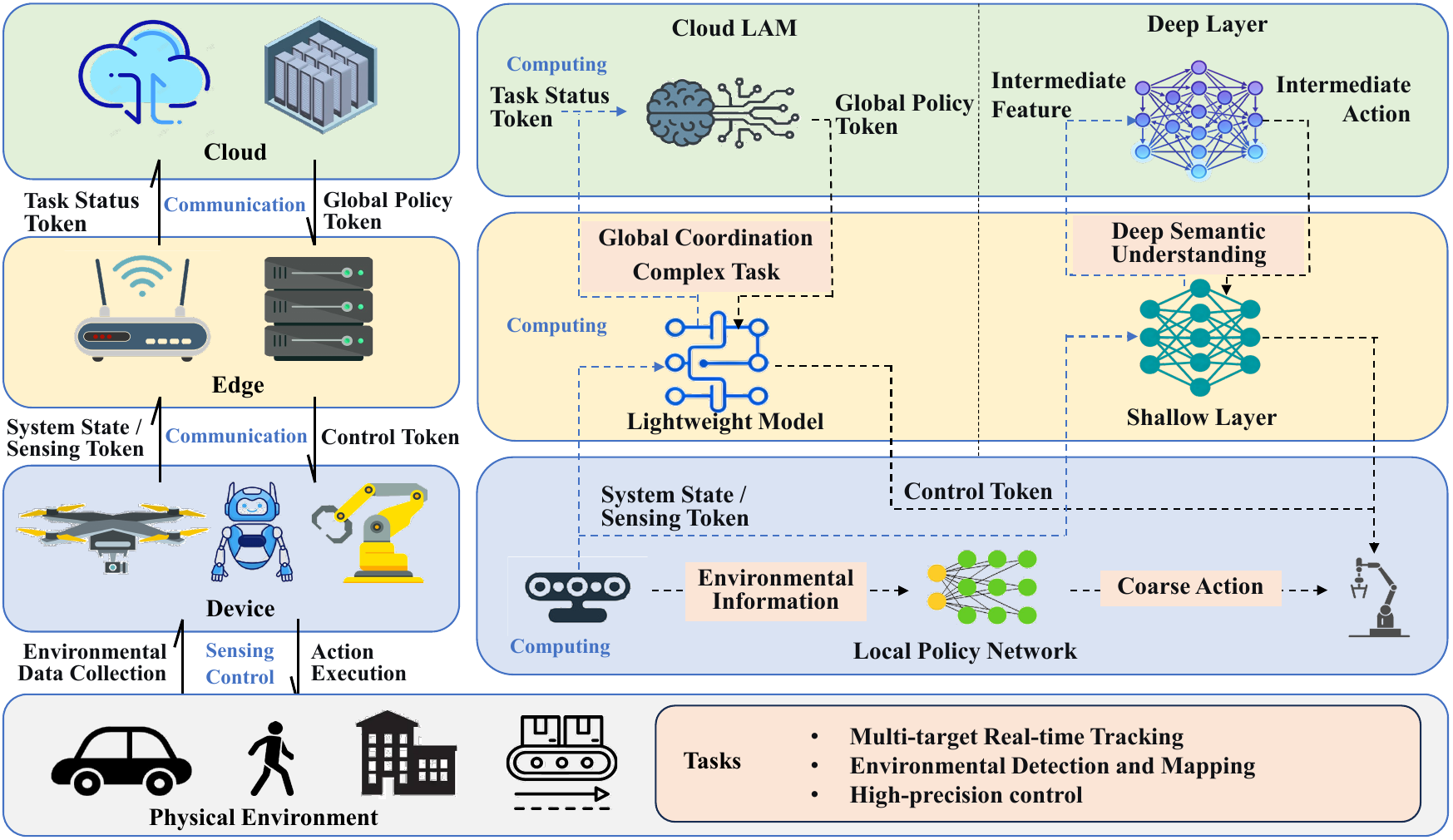}
    \caption{Cloud-edge-device deployment architectures for tokenized LAM.}
    \label{fig:deploy_architectures}
\end{figure*}

Deploying LAM in MEAN scenarios requires careful consideration of the deployment architecture. Currently, there are two mainstream cloud-edge-device collaborative deployment architectures. The first is the large-small model collaboration architecture, which deploys lightweight small models on the edge device to handle routine tasks, while the cloud-based large model serves as an enhanced processor for complex samples. The second is the model partitioning collaboration architecture, which partitions LAM hierarchically, deploying shallow layers on the edge/device and deep layers on the cloud, transmitting intermediate activations rather than raw data between nodes. As shown in Fig.~\ref{fig:deploy_architectures}.

However, the above architectures are primarily designed for general edge computing scenarios and face new challenges when directly migrated to MEAN. The core characteristics of MEAN lie in the continuous movement of agents leading to dynamic network topology changes, rigid constraints on inference latency imposed by physical closed-loop control, and the deep integration of communication, perception, computation, and control. These characteristics require specially designed deployment architectures for LAM in MEAN.

\subsubsection{Large-Small Model Collaboration Architecture}

Cloud-based large models are mainly responsible for complex reasoning, planning, and knowledge-intensive tasks, where broad generalization, deep semantic understanding, and cross-domain knowledge are essential. Edge-based small models, by contrast, handle rapid processing of routine requests, precise computation, and pre-processing near the data source, offering low-latency responses, personalization, and local data processing. This complementary division of labor is the foundation of large-small model collaboration: rather than forcing a single model to satisfy all requirements, the system assigns each part of the workload to the tier best suited for it.

Through collaborative inference and training, the two tiers achieve dynamic task allocation, task partitioning, and result fusion under the premise of preserving data locality. A typical allocation mechanism uses the small model as the primary processor for routine requests and triggers a cloud-based large model call when the small model's output confidence falls below a preset threshold. For task partitioning, a single task can be decomposed into subtasks, with the small model handling precise computation or pre-processing and the large model handling complex reasoning and planning. Hybrid mechanisms further integrate task-level and token-level collaboration—for example, speculative decoding in which the small model rapidly generates drafts and the large model verifies them in parallel, optionally combined with retrieval-augmented generation to enhance the small model's domain knowledge coverage. These mechanisms enable efficient, scalable, and reliable intelligent services in heterogeneous, resource-constrained environments while preserving data locality \cite{li2025collaborative, chen2025survey}.

\subsubsection{Model Partitioning and Distributed Inference}

The core idea of the model partitioning collaboration architecture is to partition LAM hierarchically or by components, deploying shallow layers on the edge or device tier and deep layers on the cloud tier, transmitting intermediate activations rather than raw data between nodes. Its design motivation lies in the fact that the parameter count of a complete LAM far exceeds the capacity of a single device, yet its computational graph contains natural partition points—shallow layers are responsible for general feature extraction, while deep layers handle semantic understanding and action generation.

Under this architecture, the most critical problem to be solved in the deployment architecture is the selection of the partition point. The partition point determines the trade off between computational load on edge nodes and the volume of data to be transmitted, which can be dynamically adjusted by sensing wireless channel conditions in real time by offloading more computation to the cloud when channel quality is good, and migrating computation to the edge when channel quality deteriorates \cite{chen2025adaptive}. Furthermore, the partition granularity can be refined from the layer level to the attention head level or even the token level to accommodate the heterogeneous computational requirements of different tasks \cite{younesi2025splitwise}.

\subsubsection{MEAN-specific Deployment Requirements}

The MEAN scenario possesses unique characteristics that distinguish it from general edge computing, imposing additional constraints and design considerations on deployment architectures.

In MEAN, agents continuously move within physical space, causing the connectivity topology between agents and edge nodes to change dynamically over time. An agent may move from the coverage area of one edge node to another, or even enter regions with poor network coverage. This requires the deployment architecture to possess service instance following migration capability. When an agent's attachment node changes, its ongoing LAM inference tasks need to be smoothly migrated to the new node. Unlike traditional session migration, the state of LAM inference tasks is more complex and information such as the historical tokens cached in the KV cache profoundly affects inference continuity. How to achieve low-overhead following migration while ensuring inference continuity is a core problem that MEAN deployment architectures must address.

Meanwhile, the inference results of LAM directly drive physical actuators, and the control latency requirements are typically rigid and non-violable. In general edge computing, latency exceeding expectations merely affects quality of service; in MEAN, however, latency violations may lead to missing control commands, thereby causing physical safety incidents such as collisions. Therefore, the deployment architecture must guarantee worst-case time predictability, rather than merely pursuing minimization of average latency. This means that the selection of partition points and resource allocation strategies must be optimized based on latency upper bounds rather than average values.

\subsection{Cross-Node Communication Mechanisms for Distributed LAM Inference}

In MEAN, the deployment of LAM does not place the entire model on a single node, but rather distributes its computational and inference tasks across cloud, edge, and device tiers for collaborative execution. This distributed deployment paradigm makes communication an inherent part of the inference process itself, where the data flow between nodes directly determines the efficiency and quality of LAM inference, which demand efficient and reliable communication mechanisms.

The deployment environment of MEAN endows its communication process with several distinctive characteristics: communication resources exhibit significant heterogeneity and dynamic variation across different nodes \cite{wang2025edge}, the inference-control closed loop imposes rigid constraints on communication latency, and meanwhile, transmitted content naturally contains differences in importance and semantic redundancy, enabling selective transmission \cite{erak2026adaptive}. These characteristics collectively shape the design principles of cross-node communication for distributed inference. The left half of Fig.~\ref{fig:deploy_mechanism} illustrates selected communication mechanisms for LAM deployment in MEAN.

\subsubsection{Node-aware Adaptive Transmission}

Nodes in MEAN that ranging from lightweight mobile terminals to high-performance edge servers—exhibit significant differences in communication capabilities, and the mobility of agents further exacerbates the dynamic fluctuations of wireless link quality. When distributed inference tasks need to be collaboratively completed across such heterogeneous and dynamic networks, fixed communication protocols cannot adapt.

Addressing this problem requires establishing a node-aware adaptive transmission protocol. When the deployment architecture supports both independent and collaborative execution of computational tasks, the system can continuously monitor real-time channel quality, available bandwidth, and other statuses between nodes, and dynamically adjust cross-node communication strategies accordingly. When channel conditions are favorable, the system can offload more computation to the cloud to reduce the load on local computing resources \cite{foko2024arpmec}. When channel quality degrades, agents can adjust their computation nodes in real time, migrating computational tasks that would otherwise require transmission over low-quality links to closer nodes, or switch to local computation to execute tasks independently and reduce reliance on remote communication.

\subsubsection{Deployment-Mode-Oriented Data Flows and Communication Requirements}

In mobile network environments, the deterioration of wireless link quality is hardly avoidable, while bandwidth fluctuation, packet loss, and delay jitter directly cause increased inference latency or loss of critical data. More importantly, the data flows generated under different deployment modes differ significantly in type and role, and treating them uniformly would waste scarce bandwidth and potentially undermine real-time performance. Therefore, these data flows must first be explicitly distinguished. In large-small model collaboration, nodes mainly exchange task requests, context, and inference results, while in model partition deployment, intermediate activations at the partition point must be transmitted across nodes, and in service migration scenarios, model parameters, KV caches, and request execution states may need to be transferred.

These data flows differ fundamentally in their communication requirements: task requests and context determine whether inference can be correctly initiated, inference results and intermediate activations affect inference continuity and the stability of the control loop, while model parameters and execution states place greater emphasis on consistency and integrity during migration. Consequently, the MEAN deployment architecture needs to design transmission and scheduling schemes tailored to different types of data flows. For example, prioritizing critical requests and results, compressing or selectively transmitting intermediate activations as needed, and ensuring consistency and recovery ability for migration states. The goal is to maintain stable operation of the inference-control closed loop with minimal semantic loss under constrained network conditions, while preserving reliable task execution even when bandwidth fluctuates.

\subsubsection{Cross-Node Transfer of Model and Inference States}

During distributed inference, the intermediate states that LAM relies upon across its layers and attention mechanisms, such as KV cache tensors and session contexts, often need to dynamically flow across different nodes as inference progresses. When a node becomes overloaded or link quality changes abruptly, the inference states originally handled by one node may need to be transferred to another node for continued processing.

The core objective of model and inference state migration is to ensure consistent transfer of parameters and states between old and new nodes while guaranteeing service continuity. This includes the migration of model parameters, the migration of inference states, and the assurance of service continuity. It requires the inference process to record critical intermediate states and adopt appropriate compression strategies for both model and state parameters to reduce migration overhead. Meanwhile, during migration execution, predictions of node resource fluctuations can be leveraged to proactively migrate model inference tasks and necessary state data to the target node, thereby minimizing switching latency and ensuring service continuity \cite{liu2024moei}, providing critical support for the long-term stable deployment of large action models.

\begin{figure*}[t]
    \centering
    \includegraphics[width=0.85\textwidth]{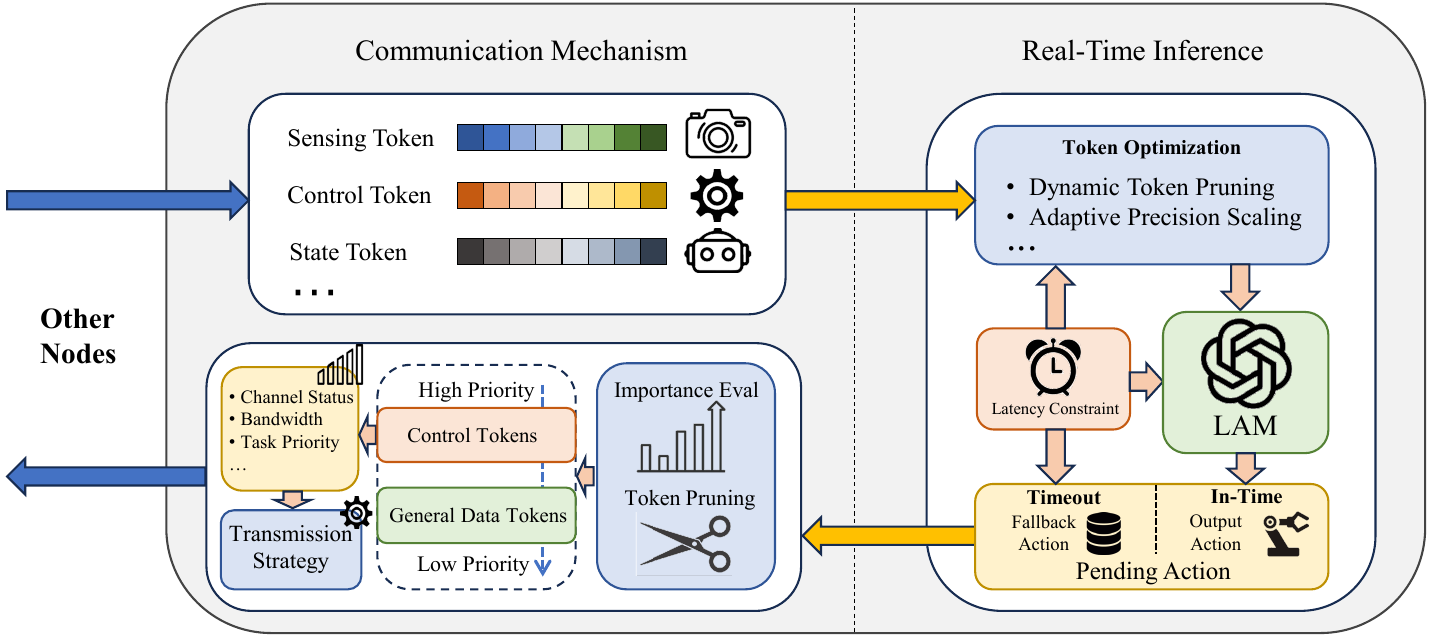}
    \caption{Communication and inference mechanisms for LAM deployment in MEAN.}
    \label{fig:deploy_mechanism}
\end{figure*}

\subsection{Real-Time Inference and Resource-Constrained Execution of LAM}

After deploying LAM across the cloud-edge-device nodes of MEAN, the execution efficiency during the inference phase directly determines whether the system can meet the real-time constraints of the physical world. More critically, LAM inference is not an isolated computational process but is embedded within a physical closed loop, which means that the action output by inference must reach the actuator within a strict time budget, otherwise the agent's state will have already changed, rendering the inference result invalid. Therefore, real-time inference mechanisms for deployment scenarios need to maximize the role of key information and reduce redundancy during reasoning, while fully considering the computing constraints of edge devices and adopting timeout handling strategies to sustain system operation when latency requirements cannot be met. The right half of Fig.~\ref{fig:deploy_mechanism} presents selected inference mechanisms for LAM deployment in MEAN.

\subsubsection{Early Exit Mechanisms and Adaptive Depth Pruning}

Early exit mechanisms place lightweight classifiers between LAM layers, each producing an action prediction and a confidence level from the current hidden state. When the confidence at an intermediate layer exceeds a preset threshold, inference terminates early at that layer and outputs the action directly, avoiding full forward propagation for every input.

This mechanism is suitable when the required inference depth varies with input complexity: simple commands in open environments may need only the first few layers, whereas precise navigation through dense obstacles still requires full forward propagation. Adaptive thresholds can further adjust exit aggressiveness according to task criticality and available time budget, lowering confidence requirements under high load or tight real-time constraints in exchange for faster response.

The key risk arises from the joint design of thresholds and inter-layer classifiers: if the threshold is set too low or confidence estimates are unreliable, complex or critical inputs may exit early before sufficient information is available, leading to incorrect actions. The resulting speedup must therefore be balanced against inference correctness.

\subsubsection{Dynamic Token Pruning and Adaptive Precision Scaling}

Different MEAN nodes—ranging from lightweight mobile robots to high-performance edge servers—exhibit orders of magnitude differences in computational capability. Deploying the same LAM across heterogeneous nodes requires a trade-off between inference accuracy and execution efficiency. The strategy of deploying multiple model variants of different sizes directly lacks flexibility and cannot adapt to dynamic changes in runtime resource states.

The core approach is to introduce runtime dynamic adjustment mechanisms that enable LAM to adapt to the capability constraints of different nodes and the demand of different tasks. And during inference, tokens that contribute little to the final output are dynamically discarded based on their attention scores or information entropy. Discarded tokens no longer participate in the computation of subsequent layers, thereby reducing sequence length and computational load \cite{tong2026flowcut}, \cite{anagnostidis2023dynamic}. The pruning ratio can be dynamically adjusted according to the current load state of the node—the higher the load, the more aggressive the pruning.

Complementing this reduction in token count, adaptive precision scaling assigns different numerical precision levels to different layers or tokens. Critical tokens or shallow embeddings maintain high precision (e.g., FP16), while non-critical tokens or deep features can be downgraded to low precision (e.g., INT4). The precision assignment scheme can be dynamically generated by a lightweight runtime analysis module that monitors the activation value distribution and numerical sensitivity of each layer, recommending the optimal precision configuration for each layer \cite{kwon2026dp}.

\subsubsection{Inference Timeout Handling}

The uncertainty of LAM inference, including varying sequence length and fluctuating network conditions leads to significant jitter in inference latency for each execution. However, the physical control loop typically requires stability in the output action frequency, which can be disrupted by latency fluctuations. More critically, when a particular inference instance severely exceeds its time budget, the agent may completely lose control input \cite{sendai2025leave}, directly threatening operational safety. Therefore, a timeout handling mechanism must be established: when LAM inference cannot complete within the current control cycle, the system should not wait but should immediately switch to an alternative action source, ensuring that the actuator always receives valid control signals.

The core of the timeout handling mechanism lies in fully utilizing local computational resources and available data to achieve effective control relaying, rather than simply stopping actions or pausing the task.

The fallback strategy for timeout handling can adopt different sources depending on the underlying deployment architecture. Under model partitioning or full deployment architectures, the intermediate layers of LAM already contain coarse-grained action predictions, when remaining time is insufficient, the best available intermediate layer output can be extracted as an emergency action, offering the advantages of strong semantic consistency and no need for additional backup model maintenance. Under the large-small model collaboration architecture, where the intermediate layer states of the cloud-based LAM are unreachable, a lightweight policy network pre-deployed on the agent's local device is required, which generates emergency actions with extremely low inference latency. The two strategies can be flexibly combined or independently selected based on the actual deployment architecture.

\subsection{Adaptive Coordination and Continuous Evolution of LAM}

After the initial deployment of LAM in MEAN, the system does not enter a static and unchanging operational state. The environment in which mobile embodied agents operate evolves continuously—surrounding objects keep changing, task objectives are constantly updated—while the system's own resource status also fluctuates dynamically, including variations in wireless channel quality and fluctuations in edge node loads. A static deployment scheme and fixed-parameter LAM model cannot adapt to such dynamic changes. Therefore, an adaptive and continuous evolution mechanism must be established, enabling deployment strategies, model parameters, and interaction semantics to co-evolve with changes in both the environment and system state \cite{lisondra2026embodied}.

\subsubsection{Dynamic Optimization of Deployment Strategy and Architecture}

The initial deployment scheme is configured based on the system state at task startup, but this state changes continuously over long-term operation. Individual edge nodes may become gradually overloaded due to an increasing number of agents being served, or an agent may enter an area with poor network coverage, degrading the connection quality to its edge node. If the deployment scheme remains unchanged, the originally optimal node allocation will evolve into a performance bottleneck.

The core approach to addressing this problem lies in establishing an online reconfiguration mechanism for deployment strategies \cite{zhang2023mobility}, \cite{ahmadpanah2025collaborative}. This mechanism continuously monitors the resource status of each node, including computational load, available memory, network round-trip latency, and battery level—along with task-level performance metrics such as end-to-end inference latency and token throughput. According to the intensity of environmental changes, when performance deviates from expectations, the system initiates dynamic adjustments of the deployment scheme, such as reallocating execution nodes for certain LAM layers or redistributing token transmission paths based on current network conditions. This ensures that the adjustment process can track the dynamic changes of both the environment and system state in real time without excessively impacting overall performance.

\subsubsection{Online Self-Adaptation of Deployed Models}

Beyond deployment strategies, pre-trained model parameters struggle to adapt to continuously changing environments and require dynamic evolution with environmental variations. Mobile embodied agents continuously accumulate new experiential data containing critical information about the current environment during task execution.

The core of online adaptation for deployed models lies in balancing adaptability and stability: the model must learn from new experiences to improve environmental adaptability while retaining its ability to handle original scenarios, preventing service interruptions or performance fluctuations caused by frequent updates. Mechanisms such as continuous learning \cite{wang2025experience}, \cite{liu2026elastic} or reinforcement learning \cite{luong2019applications} can achieve rapid environmental adaptation and long-term stable operation of deployed models through lightweight parameter updates and low-latency inference scheduling, thereby ensuring real-time responsiveness while providing sustained performance assurance for the real-world deployment of LAM.

\section{Interplay Between Emerging Technologies and MEAN}
\subsection{NTN-Enabled MEAN}
\begin{figure*}[ht]
\centering
\includegraphics[width=0.9\textwidth]{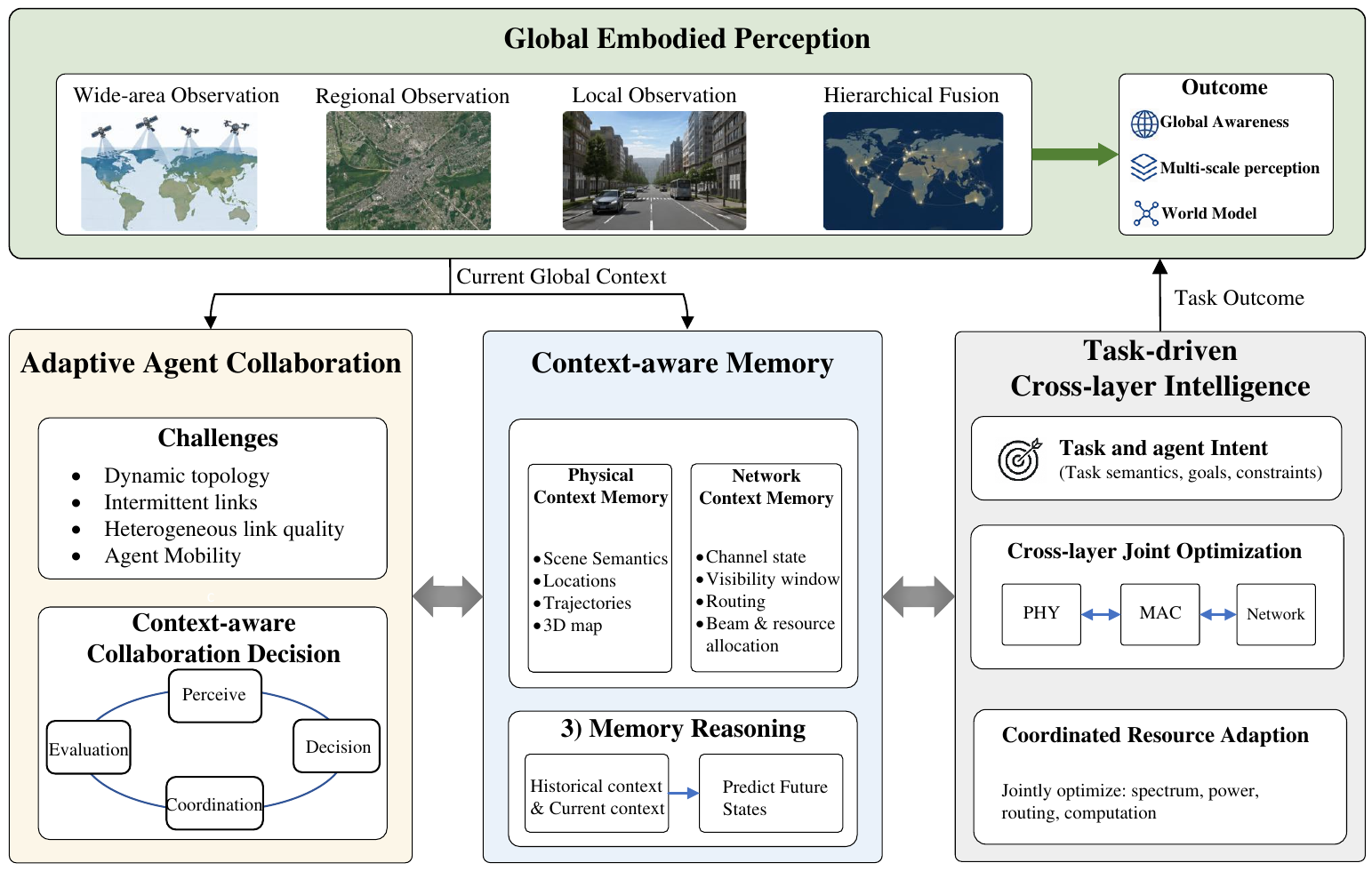}
\caption{A unified NTN-enabled MEAN framework}
\label{fig:NTN-architecture}
\end{figure*}
Embodied intelligence applications requiring collaboration beyond terrestrial coverage can benefit from non-terrestrial networks (NTNs), which provide wide-area connectivity and multi-scale observations. However, dynamic topology and intermittent connectivity complicate timely information exchange and coordinated task execution. Addressing these challenges requires NTN resource management to account for agent states, task requirements, and execution feedback. 

The NTN-enabled MEAN framework in Fig.~\ref{fig:NTN-architecture} addresses the difficulty of supporting timely embodied task execution under intermittent connectivity and delayed observations. Global embodied perception must account for variations in spatial resolution and information freshness, as intermittent links and transmission delays can render remote observations outdated before they inform local actions. Context-aware memory supports prediction of environmental changes and future contact opportunities while accounting for uncertainty in stored information. These predictions guide adaptive agent collaboration by identifying agents whose capabilities meet task requirements and whose communication opportunities permit the necessary exchanges within task deadlines. Task-driven cross-layer intelligence jointly optimizes agent selection, routing, and computing allocation to improve task performance under communication delay and interruption constraints. Execution feedback guides subsequent adjustments based on observed task outcomes.

\subsection{World Model Empowered MEAN}
\begin{figure}[!b]
\centering
\includegraphics[width=0.48\textwidth]{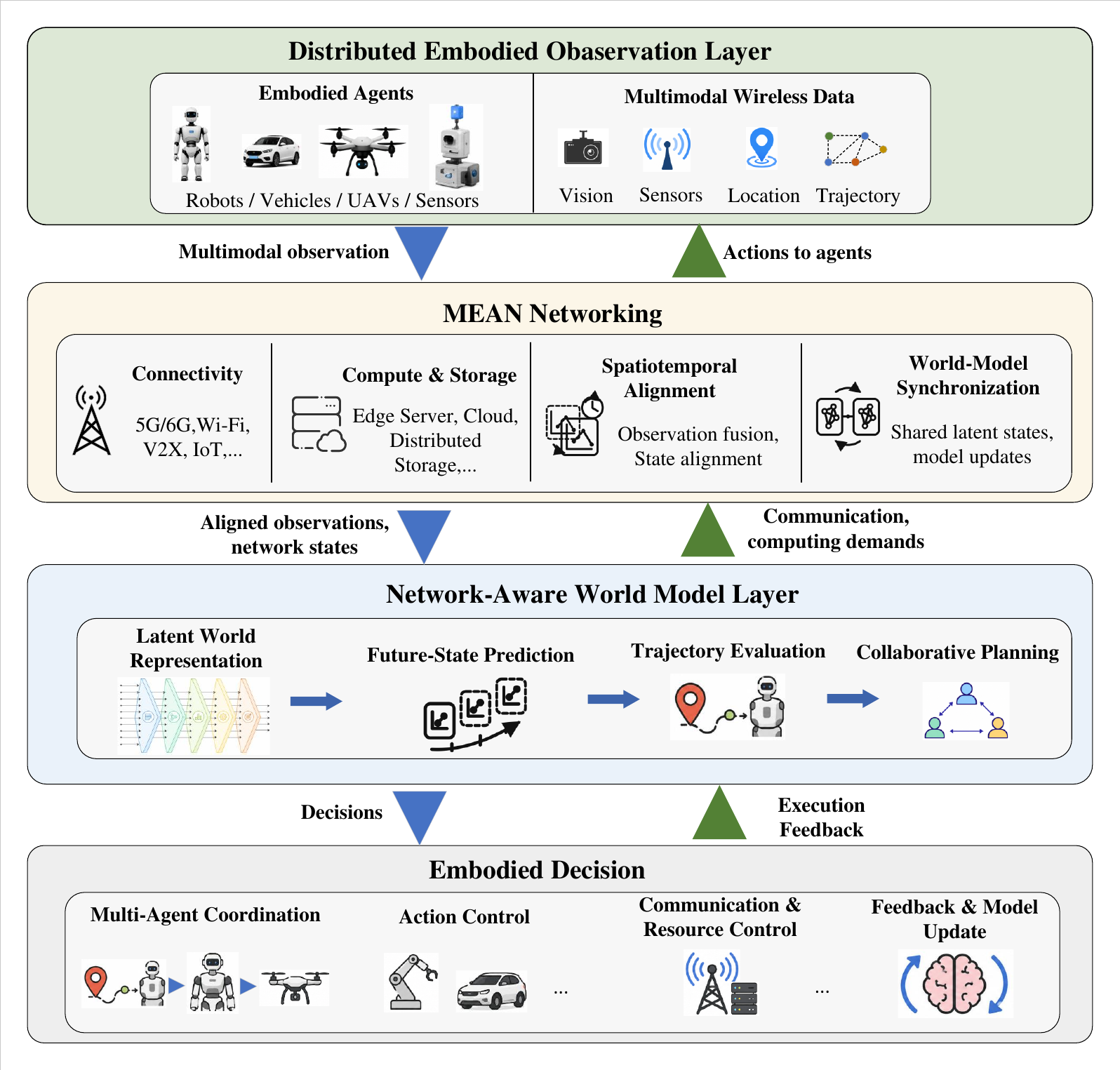}
\caption{A unified world model empowered MEAN architecture.}
\label{fig:LM-architecture}
\end{figure}

World models enable embodied agents to represent dynamic environments, predict future states, and evaluate candidate actions before execution. In MEAN, these capabilities help mitigate the limitations of partial observability and support coordination under changing wireless conditions. Fig.~\ref{fig:LM-architecture} illustrates a four-layer framework in which distributed observations and networked computing resources support world-model construction and updating, while model predictions guide task planning and resource coordination in MEAN.

The MEAN networking layer supports distributed world-model construction by sharing observations and coordinating edge–cloud processing. Spatiotemporal alignment make observations from different agents comparable, while selective exchange of latent states and model updates helps maintain consistent context with limited communication overhead.

The world-model layer learns how the environment and agents evolve in response to actions, with network conditions incorporated into its state representation. By simulating candidate action sequences, it estimates their effects on task progress and resource demand. Planning can then account for predicted outcomes and uncertainty before actions are executed.

The decision layer translates model predictions into coordinated agent actions and resource allocations. Execution feedback enables agents to adjust subsequent decisions, while discrepancies between predicted and observed outcomes guide state correction and model updating. The update frequency and planning horizon must account for communication costs and prediction errors.
At the observation layer, embodied agents and sensors collect multimodal information about the environment, tasks, and agent states.

\begin{figure*}[!t]   
\centering
\includegraphics[width=0.90\textwidth]{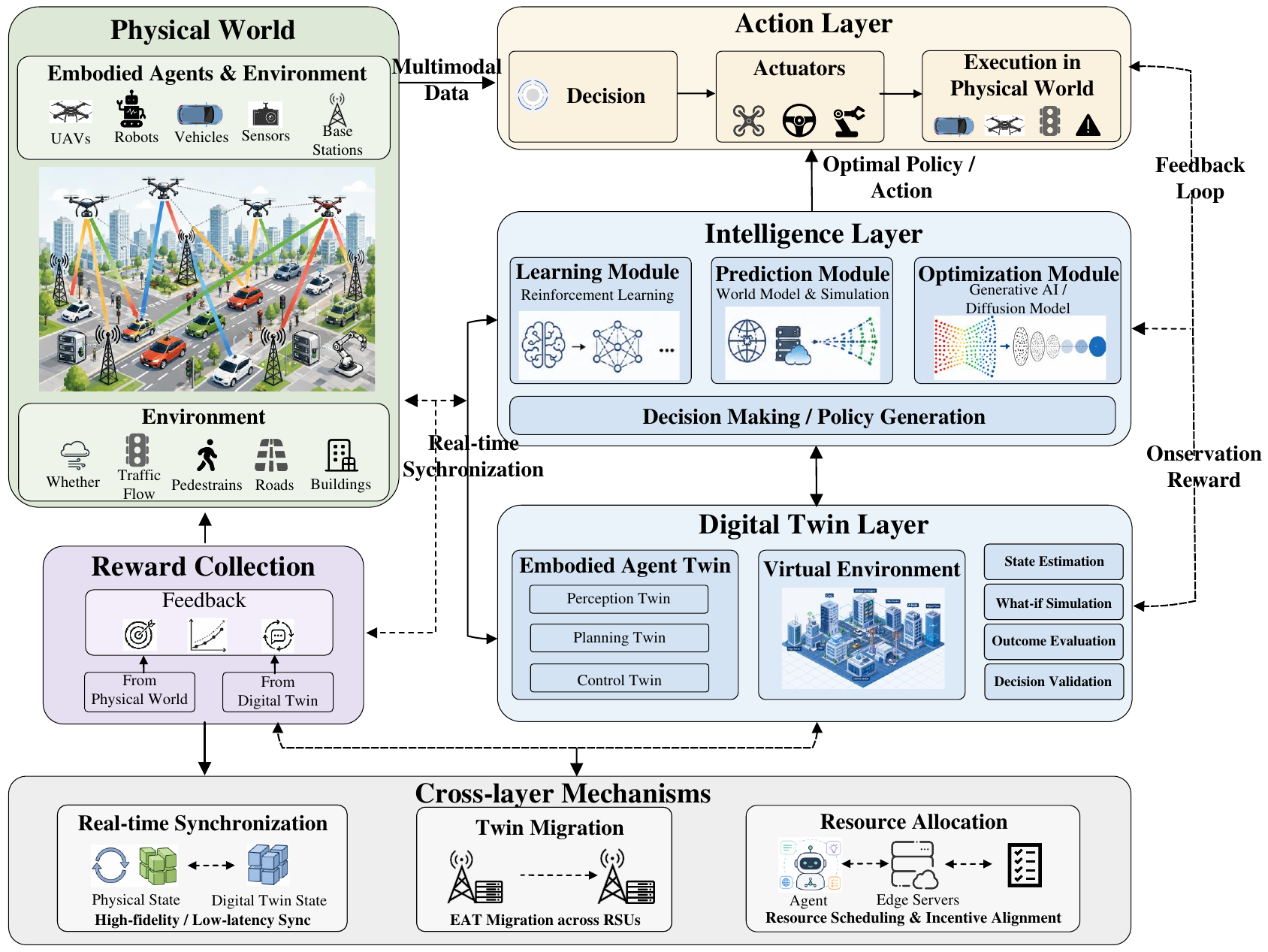}
\caption{Digital twin enabled MEAN architecture.}
\label{fig:DT-architecture}
\end{figure*}

\subsection{Digital Twin Supported MEAN}

Digital twins maintain virtual representations of embodied systems that are updated to reflect physical changes, supporting policy testing before real-world execution. Their effectiveness depends on timely updates, which can impose substantial communication and computing costs. Fig.~\ref{fig:DT-architecture} illustrates how MEAN supports these updates and connects virtual policy evaluation with physical execution through the digital twin, intelligence, and action layers.

In the physical world, embodied agents and their surroundings generate multimodal observations that provide the basis for constructing and updating digital twins.

The digital twin layer maintains a virtual representation of the embodied system and updates it as the environment and agent behaviors change. Real-time synchronization helps preserve consistency between physical and virtual states, but its frequency must balance model fidelity against communication overhead. Twin migration and resource allocation support this process as agents move and computing demands change.

The intelligence layer uses the synchronized twins to develop and evaluate policies for embodied tasks. Learning and optimization are informed by simulated outcomes and feedback from physical execution.

The action layer applies the resulting policies to embodied agents. Execution feedback updates the twins and refines subsequent decisions, closing the loop between virtual evaluation and physical operation.

\subsection{ISAC Enabled MEAN}
\begin{figure*}[t]   
\centering
\includegraphics[width=0.90\textwidth]{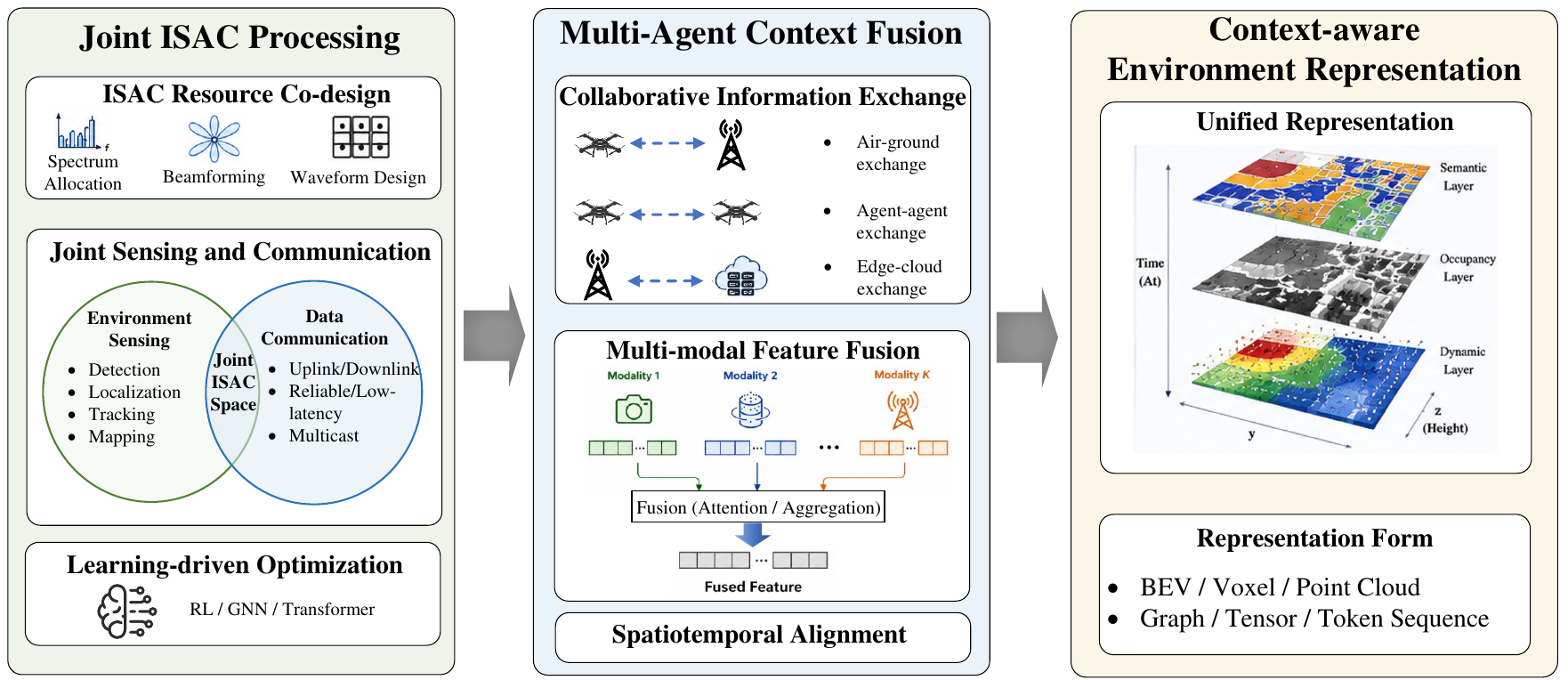}
\caption{ISAC-enabled MEAN architecture.}
\label{fig:ISAC-architecture}
\end{figure*}

Unlike conventional communication systems, where sensing and communication are designed independently, Integrated sensing and communication (ISAC) allows embodied agents to simultaneously acquire environmental information and exchange task-related data through shared wireless resources. This capability is particularly important for MEAN, where perception continuously interacts with communication, reasoning, and control during closed-loop task execution. However, realizing ISAC in MEAN remains challenging due to dynamic environments, heterogeneous sensing modalities, limited wireless resources, and the need for real-time coordination among distributed embodied agents. Therefore, ISAC should be integrated with task-oriented perception, semantic interaction, and adaptive decision-making within a unified MEAN architecture.

Fig.~\ref{fig:ISAC-architecture} illustrates the unified framework of multi-agent context fusion in ISAC-enabled MEAN. Rather than treating sensing observations from different agents independently, the proposed framework enables distributed embodied agents to collaboratively construct a unified contextual understanding of dynamic environments through collaborative information exchange, multi-modal feature fusion, and spatiotemporal alignment.

First, distributed embodied agents continuously exchange task-relevant information through heterogeneous communication links, including air–ground, agent–agent, and edge–cloud interactions. Instead of transmitting all raw sensing data, only task-oriented observations and contextual information are shared among collaborative agents, significantly reducing communication overhead while extending the perception range beyond individual sensing capabilities.

Second, heterogeneous observations collected from multiple sensing modalities are integrated through a lightweight multi-modal feature fusion module. Features extracted from different modalities are projected into a common latent space and aggregated through attention-based fusion mechanisms, allowing complementary information from visual, radio, and onboard sensing to be effectively combined into a unified feature representation.

Finally, since observations originate from distributed mobile agents operating under different viewpoints and sampling times, spatiotemporal alignment is performed before they are used for environment understanding. Through the interaction of these three components, the proposed context fusion framework enables MEAN to transform observations into a consistent context-aware environment representation, supporting coordinated decisions. 

\subsection{MEAN for Edge Intelligence}

\begin{figure}[!b]   
\centering
\includegraphics[width=0.48\textwidth]{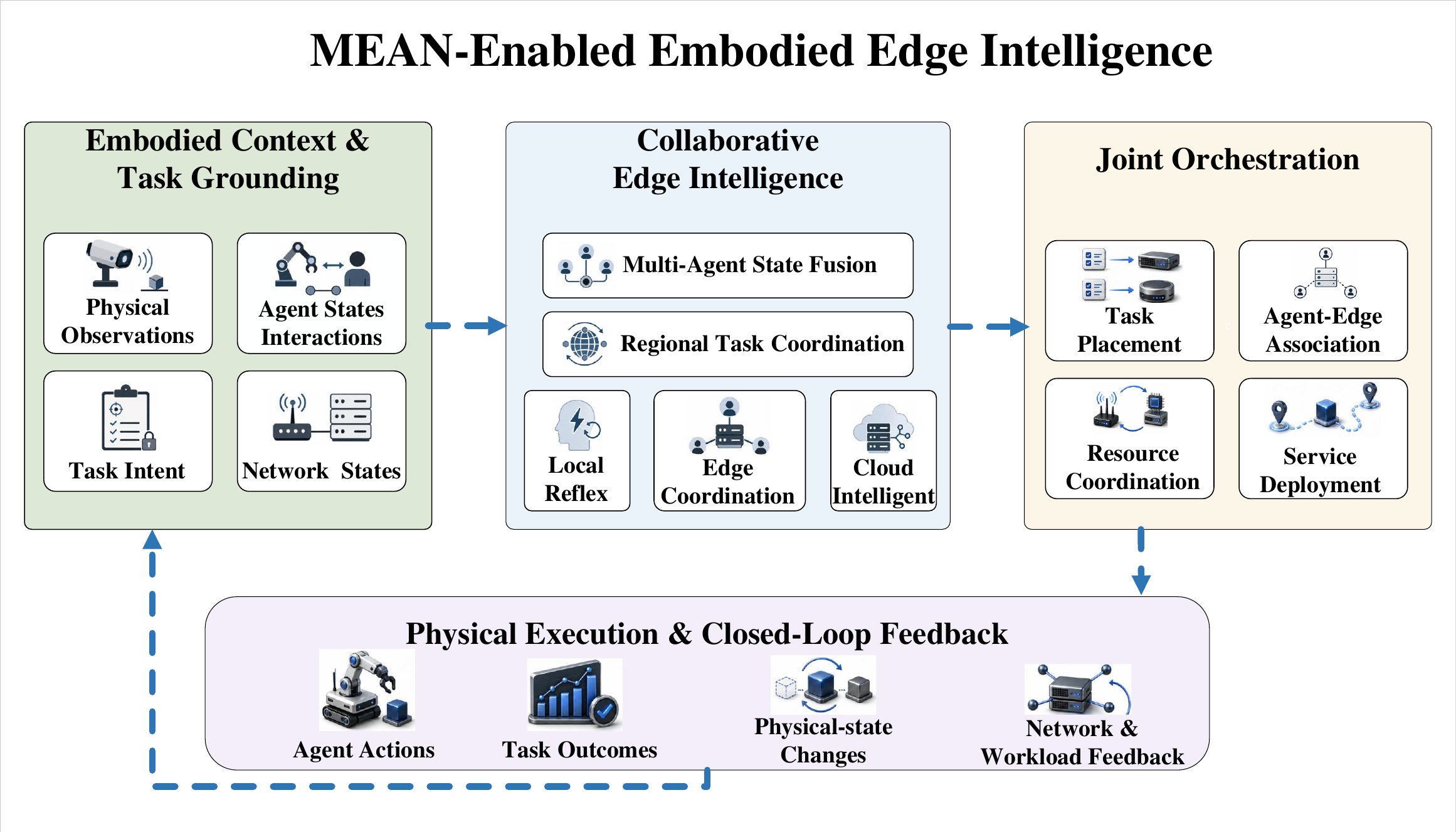}
\caption{MEAN-enabled embodied edge intelligence architecture.}
\label{fig:edge intelligence-architecture}
\end{figure}

Resource decisions in conventional edge intelligence primarily rely on network conditions and computing workloads, with limited consideration of physical environments and their implications for task execution. Edge intelligence provides MEAN with nearby computing resources and regional coordination, while MEAN incorporates embodied observations, task intentions, and execution feedback into edge decisions, enabling computing services and resource allocation to adapt to environmental changes and task objectives.

As illustrated in Fig.~\ref{fig:edge intelligence-architecture}, this interaction is realized through a device–edge–cloud architecture. Embodied agents provide local observations and task states, edge nodes fuse information from multiple agents into a shared representation of the regional environment and tasks, and the cloud supports global knowledge and long-term optimization. Edge-side fusion turns partial observations into a shared understanding of task progress and collaboration requirements, supporting multi-agent coordination. Observations and execution feedback flow upward, while task objectives and coordination policies flow downward, connecting perception and decision-making across the hierarchy.

Based on this shared context, edge nodes jointly adapt task placement, agent–edge association, and communication and computing resource allocation according to task dependencies, agent states, and environmental changes. For example, when one agent's processing result is required for others to proceed, resource allocation must account for its impact on overall task progress. Edge orchestration thus extends its objectives from computing and communication efficiency to embodied task performance. Task outcomes and environmental changes are fed back to edge nodes to update the shared context and refine subsequent orchestration, closing the loop between environmental perception, resource decisions, and physical execution.

\section{Open Problems and Future Direction}

Despite recent advances, existing technologies still lack a unified design and systematic validation framework for end-to-end embodied task execution in MEAN. This section discusses the key open problems and future research directions toward scalable, reliable, and task-oriented closed-loop embodied intelligence.

\begin{itemize}

\item \textbf{Unified theoretical foundations for MEAN:}
Existing studies often analyze communication, computation, learning, and control separately, lacking a unified theoretical framework for task-oriented embodied intelligence \cite{wen2024iscc,zhang2026edgegeneral,yang2023semantic}. The key challenge is to establish explicit relationships among network dynamics, semantic information, physical interaction, and task outcomes, since improvements in individual components do not necessarily translate into better closed-loop performance. Future research should develop unified mathematical models and analytical methods that jointly characterize task success, resource efficiency, stability, and reliability. Such foundations could clarify fundamental performance limits and provide theoretical guidance for cross-layer system optimization.

\item \textbf{Cross-layer co-design under dynamic environments:}
Existing resource management methods often optimize communication, computation, and control independently, making them difficult to adapt to dynamic embodied environments \cite{mach2017mec,chen2019drl,luong2019drl,wen2024iscc}. The key challenge lies in coordinating heterogeneous resources and task requirements that evolve across different time scales, where local optimization may compromise overall task performance. Future research should investigate task-aware cross-layer orchestration and adaptive optimization mechanisms that jointly adjust communication, computation, perception, and control policies. Particular attention should be given to maintaining closed-loop responsiveness and stability under mobility, resource fluctuations, and environmental uncertainty.

\item \textbf{Reliable and efficient token-mediated interaction:}
Existing token communication methods lack systematic mechanisms for preserving task-relevant semantics across heterogeneous embodied agents and dynamic wireless links \cite{yang2023semantic,lu2024semantics,liu2024semantic,zitkovich2023rt2}. The fundamental tradeoff is between communication efficiency and semantic fidelity, since aggressive token compression may discard information essential for coordination and reliable action execution. Future research should explore task-aware token generation, adaptive compression, semantic alignment, and context synchronization under bandwidth and reliability constraints. Developing interoperable token representations and error-resilient exchange mechanisms is also essential for maintaining consistent task understanding across agents with different sensing modalities and intelligence models.

\item \textbf{Scalable multi-agent coordination and cooperation:}
Existing multi-agent coordination approaches face scalability limitations as agent populations, task dependencies, and network interactions increase \cite{busoniu2008marl,ning2024marl,savva2019habitat,yang2025embodiedbench}. The central challenge is balancing coordination effectiveness against growing communication overhead, synchronization requirements, and distributed decision-making complexity, particularly under partial observability and dynamic network topology. Future research should investigate hierarchical coordination, distributed task allocation, and communication-efficient collaborative learning to support large-scale embodied systems. Robust mechanisms for role reassignment, local autonomy, and failure recovery are also needed to maintain cooperative task execution when connectivity or agent availability changes.

\item \textbf{Real-time deployment of large AI models in MEAN:}
Existing large-model deployment approaches remain constrained by substantial computational, memory, and energy requirements, limiting their applicability to real-time embodied systems \cite{driess2023palme,zitkovich2023rt2,kim2024llmrobotics}. The key challenge is balancing reasoning capability against inference latency, communication overhead, and control reliability, since delayed or incorrect model outputs may directly affect physical execution. Future research should investigate lightweight architectures, adaptive inference, device--edge collaboration, and uncertainty-aware execution mechanisms to achieve efficient and reliable closed-loop intelligence. These approaches must jointly consider model accuracy, resource consumption, and physical task requirements rather than optimizing inference performance alone.

\begin{figure*}[t]
    \centering
    \includegraphics[width=0.9\textwidth]{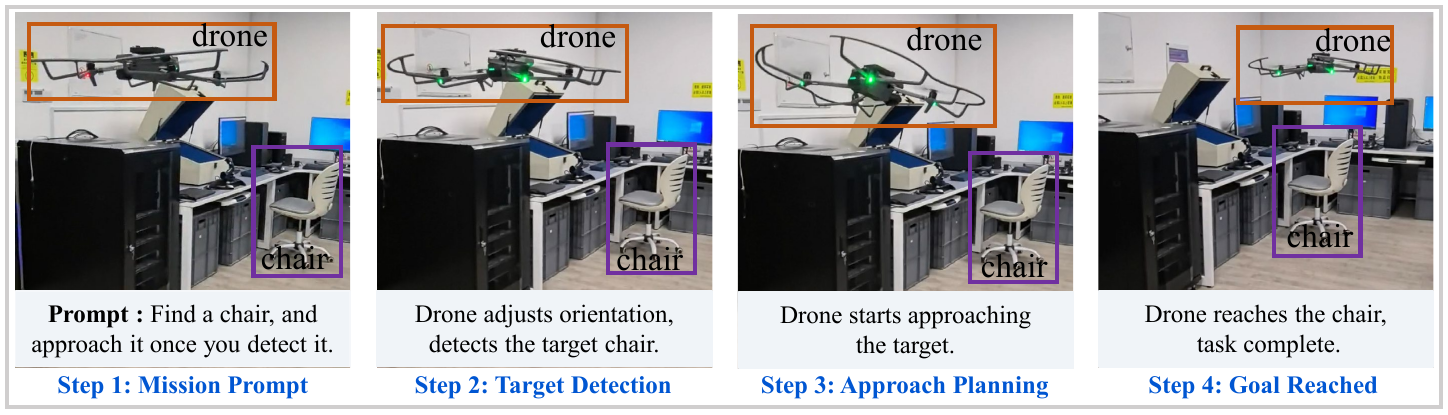}
    \caption{Representative prototype example of closed-loop embodied task execution in an Agentic AI-RAN system~\cite{sun2026agentic}.}
    \label{fig:airan_uav_case}
\end{figure*}

Fig.~\ref{fig:airan_uav_case} illustrates an Agentic AI-RAN prototype in which a UAV interprets a natural-language instruction, detects a target chair, and approaches it through wireless control. This example highlights the tight coupling among semantic reasoning, wireless communication, real-time inference, and physical action.

\item \textbf{Safety, robustness, and trustworthiness:}
Existing approaches provide limited system-level safety guarantees when learning-based decisions, wireless communication, and physical control are tightly coupled \cite{garcia2015saferl,ahn2023saycan,lu2024semantics}. The main challenge is that errors in perception, reasoning, or communication can propagate through the closed loop, potentially causing unsafe actions or task failures. Moreover, adaptive policies and large-model outputs may introduce additional uncertainty and unpredictable behavior. Future research should develop uncertainty-aware decision-making, runtime safety monitoring, fault-tolerant control, and fallback mechanisms. Integrating formal safety constraints with adaptive intelligence is essential for reliable operation under dynamic and uncertain conditions.

\item \textbf{Benchmarking and evaluation methodology:}
Existing benchmarks primarily evaluate communication, robotics, or AI capabilities separately, lacking standardized methodologies for assessing complete MEAN systems \cite{savva2019habitat,kolve2017ai2thor,shridhar2020alfworld,srivastava2022behavior,yang2025embodiedbench}. The key challenge is establishing reproducible evaluations that capture interactions among network conditions, computational resources, intelligent decisions, and physical task outcomes. Improvements in isolated metrics may not reflect actual gains in closed-loop execution. Future research should develop integrated benchmarks and experimental testbeds incorporating realistic wireless dynamics, heterogeneous agents, and multi-agent tasks. Unified evaluation protocols should jointly measure task success, latency, resource efficiency, reliability, and safety to enable meaningful system-level comparisons.

\item \textbf{Interoperability and system standardization:}
Existing embodied intelligence systems employ heterogeneous hardware, sensing modalities, control interfaces, and communication protocols, limiting interoperability and system integration \cite{liu2019edge,khan2022digitaltwin,kaushik2023isacstandardization,wen2024iscc}. The key challenge is enabling consistent task coordination and information exchange across diverse agents, intelligence models, and edge--cloud infrastructures without sacrificing platform-specific capabilities. Future research should establish common abstractions for agent capabilities, task descriptions, resource states, and semantic tokens, together with standardized interfaces for cross-layer interaction. Modular architectures and interoperable protocols will be essential for transforming isolated MEAN prototypes into scalable, reusable, and extensible systems.

\end{itemize}

\section{Conclusions}

This survey shows that supporting mobile embodied agents requires more than adding isolated AI functions to wireless networks. The central challenge is to coordinate communication, intelligence, and physical action around evolving task objectives and execution feedback. Accordingly, RL, token communication, and large AI models should not be viewed as independent technical modules, but as complementary mechanisms that jointly enable adaptive, semantically aligned, and cognitively capable operation. The future development of MEAN therefore calls for system design and evaluation to move beyond individual network or model metrics and focus on the effectiveness of the entire embodied task loop. This task-centered perspective provides a basis for integrating emerging technologies into practical networked embodied intelligence.

\balance
\bibliographystyle{IEEEtran}
\bibliography{IEEEabrv,ref}
\end{CJK}
\end{document}